\documentclass[10pt,aps,pre,twocolumn,groupedaddress,floatfix,longbibliography]{revtex4-1}

\usepackage{graphicx}
\usepackage{amsmath}
\usepackage{amssymb}
\usepackage{algcompatible}
\usepackage{hyperref}
\begin{document}

\title[Diffusion-driven self-assembly of rod-like particles]{Diffusion-driven self-assembly of rod-like particles: Monte Carlo simulation on a square lattice}

\author{Nikolai I. Lebovka}
\email[Corresponding author: ]{lebovka@gmail.com}
\affiliation{Department of Physical Chemistry of Disperse Minerals, F. D. Ovcharenko Institute of Biocolloidal Chemistry, NAS of Ukraine, Kiev, Ukraine, 03142}
\affiliation{Department of Physics, Taras Shevchenko Kiev National University, Kiev, Ukraine, 01033}
\author{Yuri Yu. Tarasevich}
\email[Corresponding author: ]{tarasevich@asu.edu.ru}
\affiliation{Laboratory of Mathematical Modeling, Astrakhan State University, Astrakhan, Russia, 414056}
\author{Volodymyr A. Gigiberiya}
\affiliation{Department of Physical Chemistry of Disperse Minerals, F. D. Ovcharenko Institute of Biocolloidal Chemistry, NAS of Ukraine, Kiev, Ukraine, 03142}
\author{Nikolai V. Vygornitskii}
\affiliation{Department of Physical Chemistry of Disperse Minerals, F. D. Ovcharenko Institute of Biocolloidal Chemistry, NAS of Ukraine, Kiev, Ukraine, 03142}
\date{\today}

\begin{abstract}
The diffusion-driven self-assembly of rod-like particles was studied by means of Monte Carlo simulation.
The rods were represented as linear $k$-mers (i.e., particles occupying $k$ adjacent sites). In the initial state, they were deposited onto a two-dimensional square lattice of size $L\times L$ up to the jamming concentration using a random sequential adsorption algorithm.
The size of the lattice, $L$, was varied from $128$ to $2048$, and periodic boundary conditions were applied along both $x$ and $y$ axes, while the length of the $k$-mers (determining the aspect ratio)  was varied from $2$ to $12$. The $k$-mers oriented along the $x$ and $y$ directions ($k_x$-mers and $k_y$-mers, respectively) were deposited equiprobably.
In the course of the simulation, the numbers of intraspecific and interspecific contacts between the same sort and between different sorts of $k$-mers, respectively, were calculated.  Both the shift ratio of the actual number of shifts along the longitudinal or transverse axes of the $k$-mers and the electrical conductivity of the system were also examined. For the initial random configuration, quite different self-organization behavior was observed for short and long $k$-mers. For long $k$-mers ($k\geq 6$), three main stages of diffusion-driven spatial segregation (self-assembly) were identified: the initial stage, reflecting destruction of the jamming state, the intermediate stage, reflecting continuous cluster coarsening and labyrinth pattern formation and the final stage, reflecting the formation of diagonal stripe domains. Additional examination of two artificially constructed initial configurations showed that this pattern  of diagonal stripe domains is an attractor, i.e., any spatial distribution of $k$-mers tends to transform into  diagonal stripes. Nevertheless, the time for relaxation to the steady state essentially increases as the lattice size growth.
\end{abstract}

\maketitle

\section{Introduction\label{sec:intro}}
Phase behavior and self-assembly in systems of elongated particles have attracted a great deal of attention for many years.
Onsager developed a density expansion theory and showed that the excluded volume effect can produce a nematic-isotropic (NI) transition in such systems~\cite{Onsager1949}.  The density of the NI transition for systems of rigid ellipsoids and cylinders has been estimated~\cite{Isihara1951}. A lattice model of holes and long rigid rods with a length-to-width ratio (aspect ratio) $k = 100$ has been analyzed~\cite{Flory1956}. Theory predicts that, below a certain concentration of holes, NI transition can occur. Similar behavior has also been observed for a system of noninteracting rods and holes for a simple cubic lattice~\cite{DiMarzio1961}. The transition predicted by Onsager's theory has been confirmed using a cluster expansion approximation for three-dimensional (3D) systems of rods oriented in three mutually perpendicular directions~\cite{Zwanzig1963}. The seventh virial coefficients of the expansion of the free energy in powers of density were calculated and it was shown that the system exhibits a van der Waals-like loop associated with the disorientation transition.

Scaled particle approaches have been tested against a fluid of rods with different aspect ratios $k$~\cite{Cotter1970,Lasher1970,Cotter1974}. Monte Carlo (MC) simulations have predicted isotropic, nematic, smectic-A, and solid phases for the 3D systems with $k\geq4$~\cite{Frenkel1988,Hentschke1992}. A direct transition from the isotropic to the smectic-A phase for a system with $k=3.2$ has been observed~\cite{Hentschke1992}. The critical density of the NI transition has been found to be inversely proportional to the aspect ratio $k$ of the rods~\cite{Isihara1951,Vroege1992}. A statistical theory for describing the packing properties and phase behavior of a granular material composed of elongated grains has also been developed~\cite{Mounfield1994}. The theory predicts that the systems need not necessarily undergo  a discontinuous first-order phase transition (even at minimum close-packing).
A kinetic approach has been applied to study the dynamic ordering of rods~\cite{Ben-Naim2006}. At
steady state, the nematic ordering was only observed at low diffusivities, whereas at large diffusivities, the system was disordered.

In two-dimensional (2D) systems, similar self-assembly  and NI transition in systems of rod-like particles have been predicted. The reduction in spatial dimensionality from 3D to 2D influences the nature of the ordered phases. For example, a perfect long-range order for 3D systems is realized whereas a quasi long-range order with algebraic decay of the order-parameter correlation function is realized for 2D systems~\cite{Frenkel1985,Bates2000}. The existence of liquid crystalline order in 2D systems has been discussed~\cite{Straley1971}. Note that the Mermin-Wagner-Berezinskii theorem forbids a long-range order in 2D systems with continuous symmetry, however, such an order may be possible for systems with the hard core interactions between the particles~\cite{Schoot1997}.

By means of MC simulation, a 2D system of hard ellipses with the aspect ratio $k=6$ has been studied~\cite{Vieillard-Baron1972}. The system exhibited two first-order phase transitions: a  solid-nematic one (at high density) and an NI one (at a density 1.5 times smaller). These transitions were attributed to geometrical factors. The density-functional approximation has been used to study the phase behavior of hard ellipses in 2D~\cite{Cuesta1989}. A continuous NI transition  was observed for the ellipses whereas there was a first-order transition for the ellipsoids. No evidence for a first order NI transition was found using MC simulation of hard rods~\cite{Frenkel1985}. The nematic phase demonstrated algebraic order (quasi long-range order) and the occurrence of a disclination-unbinding transition of the Kosterlitz-Thouless (KT) type has been suggested. MC simulations for hard ellipses with aspect ratios $k =2, 4$, and 6 have been reported~\cite{Cuesta1990}. The NI transition was only observed for  $k>2$ while the transition was first-order for $k =4$ and continuous (via disclination unbinding) for $k =6$. MC simulations of the gravitational pouring of elliptical particles with different aspect ratios $k$ ($k \in [1, 8]$) in 2D have been studied~\cite{Buchalter1992,Buchalter1992a}. For $k\geq 2$, ``amorphous'' packings with no translational order were observed. However, the ellipses formed the packings with long-range orientational order, and the ordering growing with increasing $k$. These states were named as the ``nematic glasses''.

Mean field model predicts nematic, columnar, and crystalline order in dense systems of parallel hard rods in 2D systems~\cite{Hentschke1992}. Scaled particle theory has been applied to 2D fluids of ellipses and rectangles~\cite{Schlacken1998}. Theory predicts the presence of an isotropic-nematic transition that is strongly dependent on the details of particle geometry. A density functional theory for the NI transition in a 2D system  of rods has been developed~\cite{Schoot1997}. The theory predicts the continuous (second order) NI transition of self-assembled rods in 2D  systems.

MC simulations have been applied to study the phase behavior of continuous 2D fluid systems with spherocylinders (tapered cylinders)~\cite{Bates2000}. At high density for long rods with high aspect ratios $k\gtrsim 7$, a 2D nematic phase of the KT type occurs. Shorter rods exhibit a melting transition to an isotropically arranged phase dominated by chains of particles which align side-by-side.

The formation of piles in the 2D packing of acrylic rods ($\approx 12$) constrained between two Plexiglas sheets has been studied both experimentally and in MC simulations~\cite{Stokely2003}. Orientational correlations with nematic ordering extending over two particle lengths were observed.

The different experimental studies for 2D systems of vertically vibrated rod-shaped particles
have been extensively discussed~\cite{Boerzsoenyi2013}. For example, various shapes (cylinders with cut tips, tapered tips, and rice like particles) with aspect ratios $k$ ranging from $4$ to $12.6$ were experimentally tested~\cite{Narayan2006}. The strictly cylindrical particles did not form ordered phases (smectic and nematic). Irrespective of the aspect ratio, a strong fourfold (tetratic) orientational order was always observed. Stacks with similarly oriented cylinders were formed in the tetratic arrangement. The range of tetratic order was shorter (when the distance was scaled by cylinder length) for long cylinders. In a similar study, stainless steel rods with aspect ratios of $k=20, 40$, and $60$ confined to 2D containers  have been analyzed~\cite{Galanis2006,Galanis2010}. At high density, the distinct patterns reflected competition between bulk nematic and boundary alignments. In a container of circular geometry, this competition produced bipolar configuration with two diametrically opposed point defects. As density increased, the patterning shifted from bipolar to a uniform alignment~\cite{Galanis2010}. The presence of large-scale collective swirl motions has been revealed in monolayers of vibrated granular rods (different rice species, mustard seeds and stainless steel rods with aspect ratios from $\approx 1$ to $\approx 8$)~\cite{Aranson2007}. It was speculated that the very strong sensitivity of swirling to the shape of the particles can be related to the formation of the tetratic structures.

In recent decades, much attention has been paid to the study of self-assembly in systems of linear $k$-mers (particles occupying $k$ adjacent adsorption sites) deposited on 2D lattices. A linear $k$-mer represents the simplest model of an elongated particle with an aspect ratio of $k$.
Computer simulations have been extensively applied to investigate percolation and jamming phenomena for the random sequential absorption (RSA) of $k$-mers~\cite{Leroyer1994PRB,Vandewalle2000,Kondrat2001PRE,Cornette2003epjb,Longone2012PRE,Tarasevich2012PRE,Budinski2016JSM,Lebovka2015PRE,Tarasevich2015JPhCS}. Various anomalies in the properties of the systems' dependence on the length of the $k$-mers have been reported. For example, the jamming concentration decreases monotonically when approaching the asymptotic value of $p_j = 0.66 \pm 0.01$ at large values of $k$. The percolation threshold $p_c$ is a non-monotonic function of the length $k$, with a minimum at a particular length of the $k$-mers ($k \approx 13$) and, presumably, percolation is impossible for very long $k$-mers ($k \gtrsim 10^4$)~\cite{Tarasevich2012PRE}.

The irreversible RSA process leads to a non-equilibrium state of the system, and further self-organization in the deposited film is possible owing to deposition-evaporation processes or the diffusion motion of the $k$-mers. Several problems related to such type of self-assembly of $k$-mers have previously been discussed~\cite{Ghosh2007,Lopez2010,Loncarevic2010,Kundu2013,Kundu2013a,Matoz-Fernandez2012,Lebovka2017}. Dynamic MC simulations using a deposition-evaporation algorithm for simulation of the dynamic equilibrium of $k$-mers on square lattices have been applied~\cite{Ghosh2007}. For long $k$-mers ($k\geq 7$),  two entropy-driven transitions as a function of density, $p$, were revealed: first, from a low-density isotropic phase to a nematic phase with an intermediate density at $p_{in}$, and, second, from the nematic phase to a high-density disordered phase at $p_{nd}$. A lattice-gas model approach has been applied to study the phase diagram of self-assembled $k$-mers on square lattices~\cite{Lopez2010}. It has been observed that the irreversible RSA process leads to an  intermediate state with purely local orientational order and, in the equilibrium model, the nematic order can be stabilized for sufficiently long $k$-mers~\cite{Matoz-Fernandez2012}. For example, for $k=7$, $p_{in}\approx 0.729$~\cite{Matoz-Fernandez2012} and $p_{nd}\approx 0.917$~\cite{Kundu2013a}.

On a square lattice, in the course of diffusion, the horizontal and vertical stacks of $k$-mers separated from one another and a characteristic coarsening was observed~\cite{Lebovka2017}. Such self-assembly of $k$-mers can significantly affect the properties of the films. The properties of electrically conductive films filled with elongated particles (e.g., carbon nanotubes) are of particular interest in the production of electrodes for super-capacitors, thin film transistors and fuel cells~\cite{Du2007,Yu2016}. Nevertheless, in spite of great interest in the problem, the impact of diffusion-driven self-assembly on the percolation and electrical conductivity of such films has  never previously been discussed in the literature.

This paper analyzes the processes of diffusion-driven  self-assembly of linear $k$-mers on a square lattice   by means of kinetic MC simulation. The initial state was produced using RSA with isotropic orientations of the $k$-mers, after which the $k$-mers underwent translation diffusion. The kinetics of the changes of structure and electrical conductivity in different directions were analyzed.

The rest of the paper is constructed as follows. In Sec.~\ref{sec:methods}, the technical details of the simulations are described, all necessary quantities are defined, and some test results are given. Section~\ref{sec:results} presents our principal findings. Section~\ref{sec:conclusion} summarizes the main results.

\section{Computational model\label{sec:methods}}

In the kinetic MC simulation, the RSA model was used to produce an initial homogeneous and isotropic distribution of linear $k$-mers (i.e., particles occupying $k$ adjacent sites) in a 2D film~\cite{Evans1993RMP}. The rod-like particles were deposited randomly and sequentially, and their overlapping with previously placed particles was forbidden.

The problem was approached using a square lattice of size $L\times L$. In the present work, almost all calculations were performed using $L=256$. In some cases, scaling analysis for lattices of sizes up to $L=2048$ was performed. Periodic boundary conditions were applied along both the $x$ and $y$ axes. The length of the $k$-mers (aspect ratio)  was varied from $2$ to $12$. Isotropic orientation of the  $k$-mers was assumed, i.e., $k$-mers oriented along the $x$ and $y$ directions ($k_x$-mers and $k_y$-mers, respectively) were equiprobably in their  deposition. This corresponded to the zero value of a mean order parameter of the system, defined as
\begin{equation}\label{eq:s}
s= \frac{n_y - n_x}{n_x + n_y},
\end{equation}
where $n_x$ and $n_y$  ($n_x=n_y$) are the numbers of $k_x$-mers and $k_y$-mers, respectively.

For a random initial configuration (r configuration), the $k_x$-mers and $k_y$-mers were randomly deposited over the entire lattice. In some cases, we divided the lattice in two areas of a specific shapes. The initial states were generated by deposition of only $k_x$-mers inside one of these areas and only $k_y$-mers inside the other area. The areas were either (a) two stripes oriented in the vertical direction (v configuration) or (b) diagonal stripes (d configuration).

The concentration of the particles corresponded to the jamming state, $p_j$. In this state, no additional $k$-mer can be placed because the presented voids are too small or of inappropriate shape. The values of $p_j$ for different values of $k$ have recently been calculated at the thermodynamic limit~\cite{Lebovka2011}, while our estimations of $p_j$ for a particular lattice size $L=256$ have also  recently been presented in~\cite{Lebovka2017}. The total number of $k$-mers at the jamming state, $N_k$, was equal to $p_jL^2/k$.

The diffusion of $k$-mers was simulated using the kinetic MC procedure (see Appendix~\ref{app:alg}). For fairly dense systems in the jamming state, rotational diffusion is impeded, especially for large values of~$k$. This is the reason why only translational diffusion was taken into consideration in our simulation.

At each step, an arbitrary $k$-mer was randomly chosen and a translational shift by one lattice unit along either the longitudinal ($\parallel$) or the transverse ($\perp$) axis of the $k$-mer was attempted. Equal probabilities of such a translational shift along the longitudinal or transverse axes of the $k$-mers were assumed. One time step of the MC computation, which corresponds to an attempted displacement of the total number of $k$-mers in the system, $N_k$, was taken as the MC time unit. Time  counting was started from the value of $t_{MC}=1$, being the initial moment (before diffusion), and the total duration of the simulation was typically $10^7$ MC time units, although in some cases even up to $10^9$.

Video~\ref{vid:Patternsu0} presents examples of the $k$-mer patterns at different moments in time, $t_{MC}$, for $k=6$ and the initial r configuration. At the initial moment, at $t_{MC}=1$, the deposited $k$-mers tend to align parallel to each other and stacks of the horizontally ($x$ stacks) and, typically,  vertically ($y$ stacks) oriented $k$-mers were observed. These stacks can be represented as squares of size $\approx k \times k$. In the jamming state,  small voids between the stacks were also present. These results fully correspond with previously reported data~\cite{Manna1991,Vandewalle2000,Lebovka2011,Lebovka2017}. Comparison of the patterns at different  times evidenced the possibility of diffusion-driven spatial segregation (self-assembly) in the systems (Video~\ref{vid:Patternsu0}). Evident separation of domains of $k_x$-mers and $k_y$-mers was observed. This phase ordering resembles the coarsening (Ostwald ripening) in phase-separating systems, such as binary alloys or in the anisotropic ferromagnet (Ising model)~\cite{Bray2002}. At late stages ($t_{MC}\geq10^6$), very large percolating domains of the horizontal and vertical stacks were formed.
%%%%%%%%%%%%%%%%%%%%%%%%%%%%%%%%%%%%%%%%%%%%%%%%%%%%%%%%%%%%%%%%%%%%%%%%%%%%%%%%%%%%%%%%%%%%%%%%%%%%%%%%%%%%%%%%%%%%%%%%%%%%%%%%%%%%%%%%%%%%%%%%%%%
\begin{video}[htbp]
\centering
  \includegraphics[width=\linewidth]{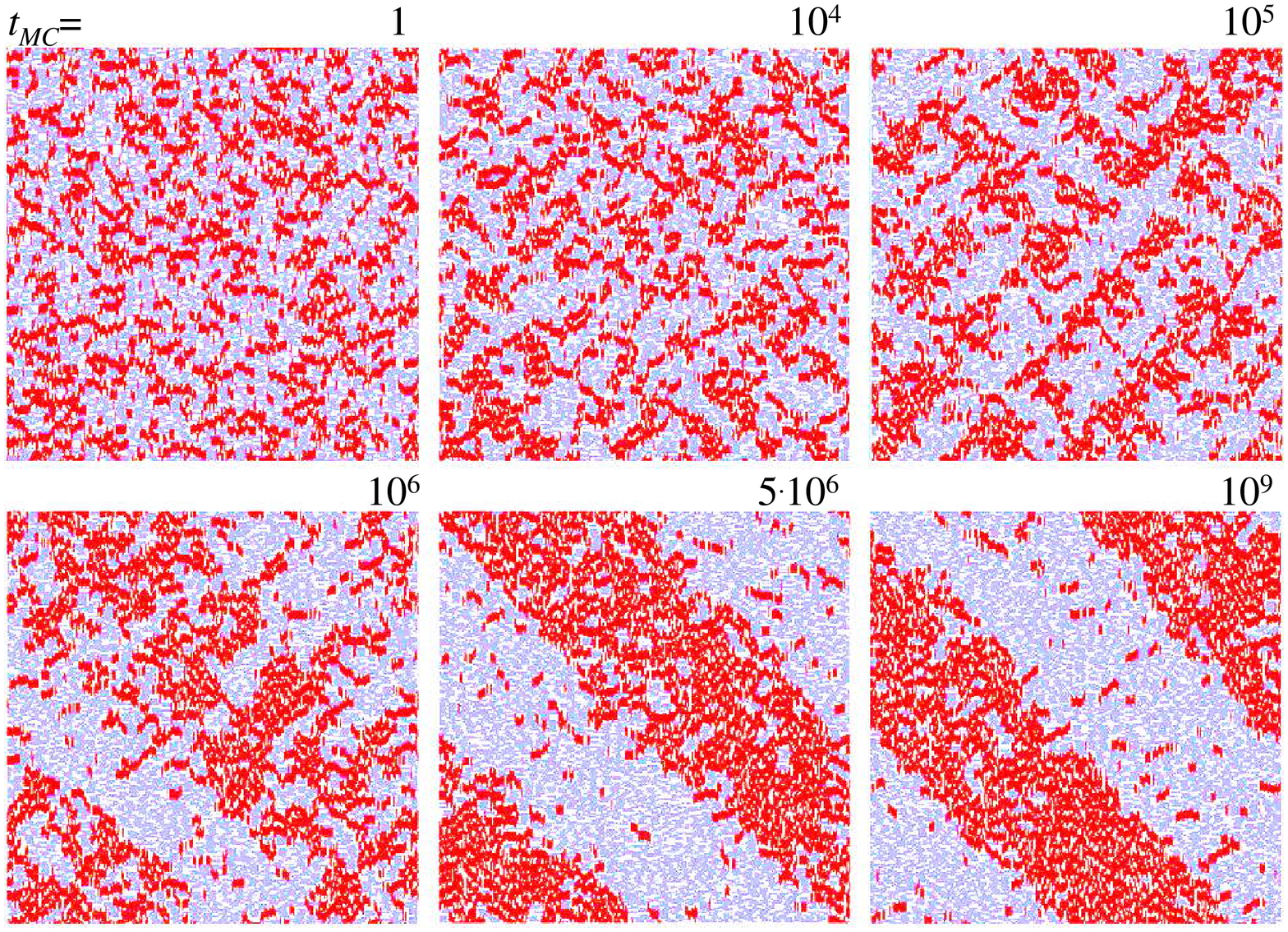}
  \setfloatlink{videok6.avi}
\caption{Examples of patterns  at different  times, $t_{MC}$. Here, $L=256$, $k=6$ and the initial r configuration were used. The concentration of $k$-mers corresponds to the jamming state, $p_j\approx 0.769$~\cite{Lebovka2017}.
\label{vid:Patternsu0}}
\end{video}
%%%%%%%%%%%%%%%%%%%%%%%%%%%%%%%%%%%%%%%%%%%%%%%%%%%%%%%%%%%%%%%%%%%%%%%%%%%%%%%%%%%%%%%%%%%%%%%%%%%%%%%%%%%%%%%%%%%%%%%%%%%%%%%%%%%%%%%%%%%%%%%%%%%

For characterization of these processes, a range of parameters were calculated during the course of the simulation:
\begin{itemize}
  \item the number of intraspecific contacts (per monomer) between the same sorts of $k$-mers (i.e., horizontal-horizontal or vertical-vertical), $n$;
  \item the number of interspecific contacts (per monomer) between the different sorts of $k$-mers (i.e., horizontal-vertical), $n_{xy}$;
  \item the shift ratio of the actual number of shifts along the longitudinal or transverse axes of the $k$-mers, $R$;
  \item the electrical conductivity.
\end{itemize}

\begin{figure}[htbp]
  \centering
  \includegraphics[width=0.9\columnwidth]{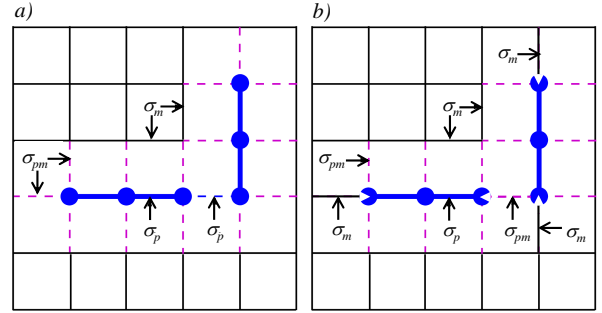}
  \caption{Fragment of a square lattice with two deposited 3-mers of different orientations. The conductivities of the bonds for (a) conventional and (b) insulating ends models are indicated.\label{fig:lattice}}
\end{figure}

Note, that the maximum value $n$ or $n_{xy}$ can attain is $2(1+1/k)$. For calculation of the electrical conductivity in the horizontal, $\sigma_x$, or vertical, $\sigma_y$, directions, the periodic boundary conditions were removed and two conducting buses were applied to the opposite borders of the lattice in the corresponding $y$ or $x$ directions. The electrical conductivities were calculated between these buses, and the mean value $\sigma=(\sigma_x +\sigma_y)/2$ was evaluated. For the conventional model, the different electrical conductivities of the bonds between empty sites, $\sigma_m$, filled sites, $\sigma_p$,  and empty and filled sites, $\sigma_{pm}=2\sigma_p \sigma_m / (\sigma_p+\sigma_m)$ were assumed (Fig.~\ref{fig:lattice}a). For the insulating ends model, the two ends of each $k$-mer were isolated and the electrical conductivities of the bonds were calculated as demonstrated in Fig.~\ref{fig:lattice}b. We put $\sigma_m =1$, and $\sigma_p= 10^6$ in arbitrary units. The Frank-Lobb algorithm was applied to evaluate the electrical conductivity~\cite{Frank1988PRB}. In the calculations, the logarithm of the effective conductivity was averaged over different runs (see~\cite{Tarasevich2016} for the details).
%%%%%%%%%%%%%%%%%%%%%%%%%%%%%%%%%%%%%%%%%%%%%%%%%%%%%%%%%%%%%%%%%%%%%%%%%%%%%%%%%%%%%%%%%%%%%%%%%%%%%%%%%%%%%%%%%%%%%%%%%%%%%%%%%%%%%%%%%%%%%%%%%%%
%%%%%%%%%%%%%%%%%%%%%%%%%%%%%%%%%%%%%%%%%%%%%%%%%%%%%%%%%%%%%%%%%%%%%%%%%%%%%%%%%%%%%%%%%%%%%%%%%%%%%%%%%%%%%%%%%%%%%%%%%%%%%%%%%%%%%%%%%%%%%%%%%%%

\begin{figure}[htbp]
  \centering
  \includegraphics[width=\columnwidth]{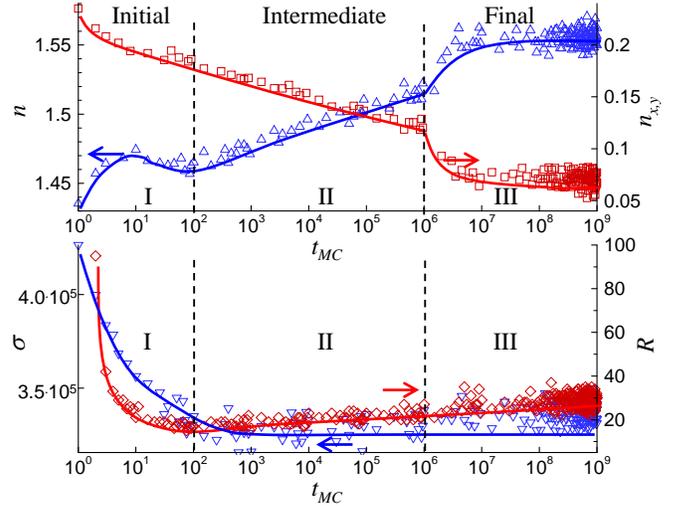}
\caption{Number of intraspecific, $n$, and interspecific, $n_{xy}$, contacts, the shift ratio, $R$, and electrical conductivity for conventional model, $\sigma$, versus the MC time $t_{MC}$. The data correspond to the same MC run as for the patterns presented in Video~\ref{vid:Patternsu0},  $L=256$, $k=6$, with use of the initial r configuration. The concentration of the $k$-mers corresponds to the jamming state, $p_j\approx 0.769$~\cite{Lebovka2017}.
\label{fig:Contacts}}
\end{figure}

Figure~\ref{fig:Contacts} presents examples of the evolution over time of the number of intraspecific, $n$, and interspecific, $n_{xy}$, contacts, the shift ratio, $R$, and the electrical conductivity for the conventional model, $\sigma$, for the same MC run that generated the for patterns presented in Video~\ref{vid:Patternsu0}.
%%%%%%%%%%%%%%%%%%%%%%%%%%%%%%%%%%%%%%%%%%%%%%%%%%%%%%%%%%%%%%%%%%%%%%%%%%%%%%%%%%%%%%%%%%%%%%%%%%%%%%%%%%%%%%%%%%%%%%%%%%%%%%%%%%%%%%%%%%%%%%%%%%%
%%%%%%%%%%%%%%%%%%%%%%%%%%%%%%%%%%%%%%%%%%%%%%%%%%%%%%%%%%%%%%%%%%%%%%%%%%%%%%%%%%%%%%%%%%%%%%%%%%%%%%%%%%%%%%%%%%%%%%%%%%%%%%%%%%%%%%%%%%%%%%%%%%%

Comparison of the data presented in Video~\ref{vid:Patternsu0} and Fig.~\ref{fig:Contacts} allows the following preliminary conclusion to be drawn.
We can identify several stages of the diffusion-driven spatial segregation (self-assembly) of the system under consideration.
\begin{description}
  \item[Initial stage] During the initial stage, the system undergoes a drastic transition, namely, the jammed state turns into a non-jammed state. Upon this transition, the number of  intraspecific contacts, $n$, goes through a maximum, while the values $n_{xy}$, $R$, and $\sigma$ decrease noticeably. The duration of the transition mode is of the order $10^2$ MC steps. Its end corresponds to the minimum of the curve $R(t_{MC})$.
  \item[Intermediate stage]
  In this stage ($t_{MC}=10^2-10^6$), continuous cluster coarsening and labyrinth pattern formation can be observed. The labyrinth structures form during the middle of the stage and then transform into the germs of the stripe domains toward its end (see Video~\ref{vid:Patternsu0}). The values of $n$ and $R$ increase and the value of $n_{xy}$ decreases near proportionally to the time $\log_{10}(t_{MC})$ (Fig.~\ref{fig:Contacts}). The end of this stage corresponds to the inflection points in the curves $n (t_{MC})$ and $n_{xy}(t_{MC})$ in Fig.~\ref{fig:Contacts}.
  \item[Final stage] During the third stage, regular patterns begin to form. These patterns appear as  striped diagonal domains oriented at $45^\circ$ to the lattice side. The stripe formation finishes at times in the order of $t_{MC}\approx 5\times10^6$. Later in the MC simulation, the system stabilizes in a non-equilibrium steady state. The diagonal stripe domains on a plane correspond to a torus divided into two equal parts (see Supplemental Material at [URL will be inserted by publisher] for a video of the torus).
\end{description}

For each given value of $k$, the computer  experiments were repeated $100$ times, and then the data were averaged. The error bars in the figures correspond to the standard error of the mean. When not shown explicitly, they are of the order of the marker size.

\section{Results and Discussion\label{sec:results}}
\subsection{Initial r configuration}
\begin{figure}[htbp]
  \centering
  \includegraphics[width=\linewidth]{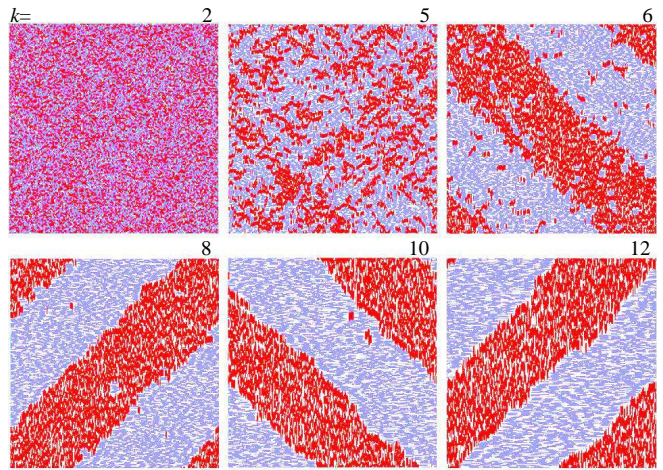}
\caption{Examples of patterns at $t_{MC}=10^7$  for different values of length of the $k$-mers.
Here, $L=256$ and the initial r configuration were used. The concentration of $k$-mers corresponds to the jamming state~\cite{Lebovka2017}.
\label{fig:All_patterns}}
\end{figure}

Figure~\ref{fig:All_patterns} presents the patterns for different values of the length of $k$-mers for  relatively long MC simulations, $t_{MC}=10^7$. The initial r configuration ($L=256$) was used. For short $k$-mers, cluster coarsening was observed. This self-assembly of rod-like particles became more evident and the clusters grew more in size with increase of $k$. Stripe domains were only observed in the case of  long rods with high aspect ratios $k\geq 6$.
%%%%%%%%%%%%%%%%%%%%%%%%%%%%%%%%%%%%%%%%%%%%%%%%%%%%%%%%%%%%%%%%%%%%%%%%%%%%%%%%%%%%%%%%%%%%%%%%%%%%%%%%%%%%%%%%%%%%%%%%%%%%%%%%%%%%%%%%%%%%%%%%%%%
%%%%%%%%%%%%%%%%%%%%%%%%%%%%%%%%%%%%%%%%%%%%%%%%%%%%%%%%%%%%%%%%%%%%%%%%%%%%%%%%%%%%%%%%%%%%%%%%%%%%%%%%%%%%%%%%%%%%%%%%%%%%%%%%%%%%%%%%%%%%%%%%%%%

\begin{figure}[htbp]
  \centering
  \includegraphics[width=0.9\columnwidth]{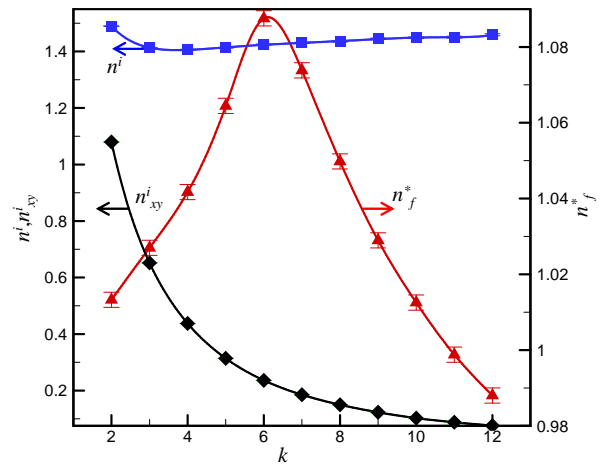}
\caption{Relationship between the  initial number (i.e., at $t_{MC}=1$) of intraspecific, $n^i$, and  interspecific, $n^i_{xy}$, contacts versus the length of the $k$-mers. The normalized numbers of intraspecific contacts, $n^*_f$ versus $k$  at  $t_{MC}=10^7$ are also presented. Here, $L=256$ and the initial r configuration were used. The concentration of $k$-mers corresponds to the jamming state~\cite{Lebovka2017}.
\label{fig:f06_contacts}}
\end{figure}

The values of the number of intraspecific, $n^i$, and  interspecific, $n^i_{xy}$, contacts for different values of $k$-mers for the initial moment $t_{MC}=1$ before diffusion began are presented in Fig.~\ref{fig:f06_contacts}. The $n^i(k)$ dependence was weak and a  minimal number of intraspecific contacts ($n^i\approx 1.407$) was observed at $k=4$. The $n^i_{xy}(k)$ dependence was fairly strong
and could be well-approximated by the following equation
\begin{equation}\label{Eq:1}
n^i_{xy}=(a/\sqrt{k} - b)^2,
\end{equation}
where $a=1.843\pm 0.016$, $b=0.2620\pm 0.009$,  and  the coefficient of determination is $r^2=0.9998$.
%%%%%%%%%%%%%%%%%%%%%%%%%%%%%%%%%%%%%%%%%%%%%%%%%%%%%%%%%%%%%%%%%%%%%%%%%%%%%%%%%%%%%%%%%%%%%%%%%%%%%%%%%%%%%%%%%%%%%%%%%%%%%%%%%%%%%%%%%%%%%%%%%%%
%%%%%%%%%%%%%%%%%%%%%%%%%%%%%%%%%%%%%%%%%%%%%%%%%%%%%%%%%%%%%%%%%%%%%%%%%%%%%%%%%%%%%%%%%%%%%%%%%%%%%%%%%%%%%%%%%%%%%%%%%%%%%%%%%%%%%%%%%%%%%%%%%%%

\begin{figure}[htbp]
  \centering
  \includegraphics[clip,width=0.9\columnwidth]{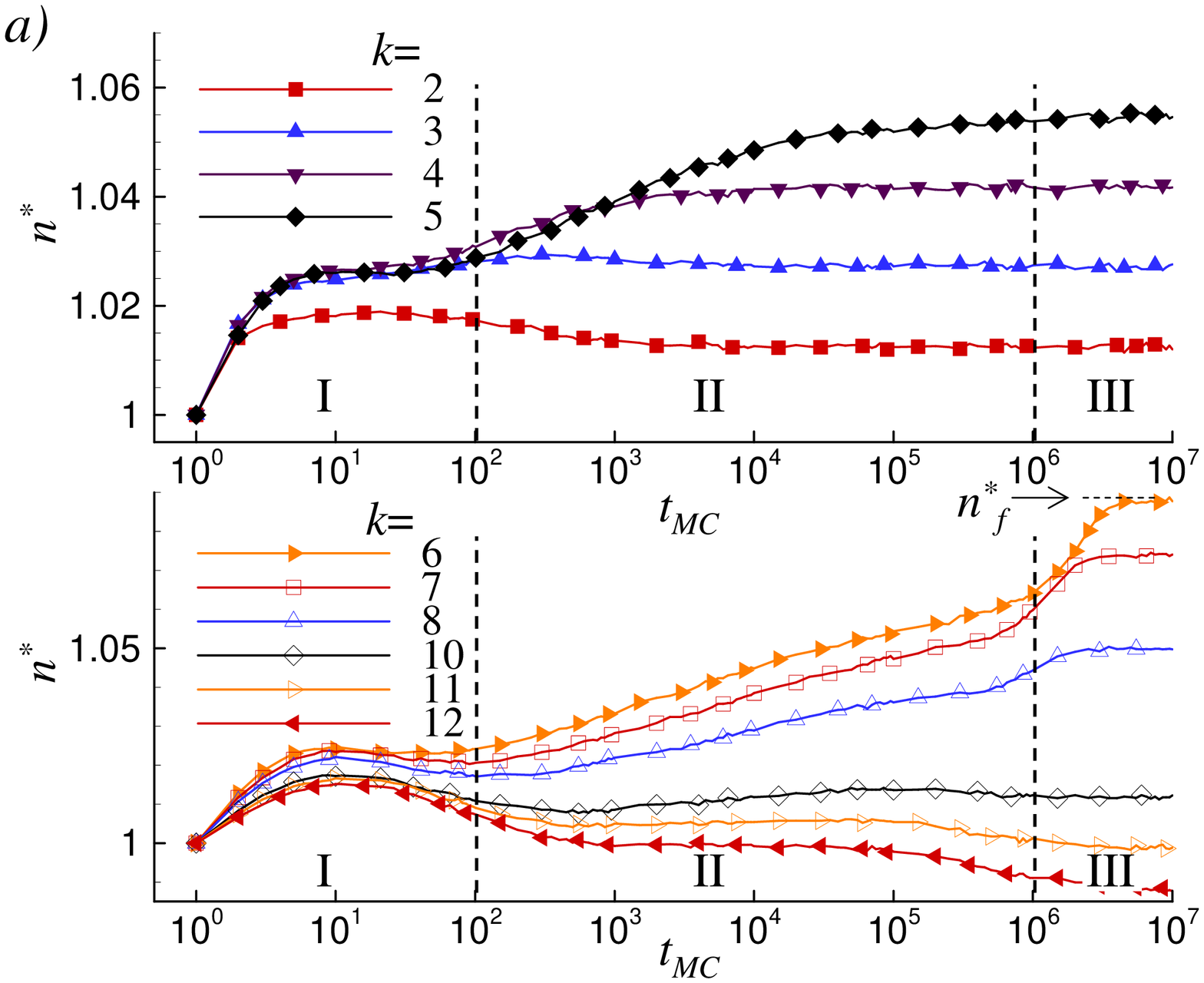}
  \includegraphics[clip,width=0.9\columnwidth]{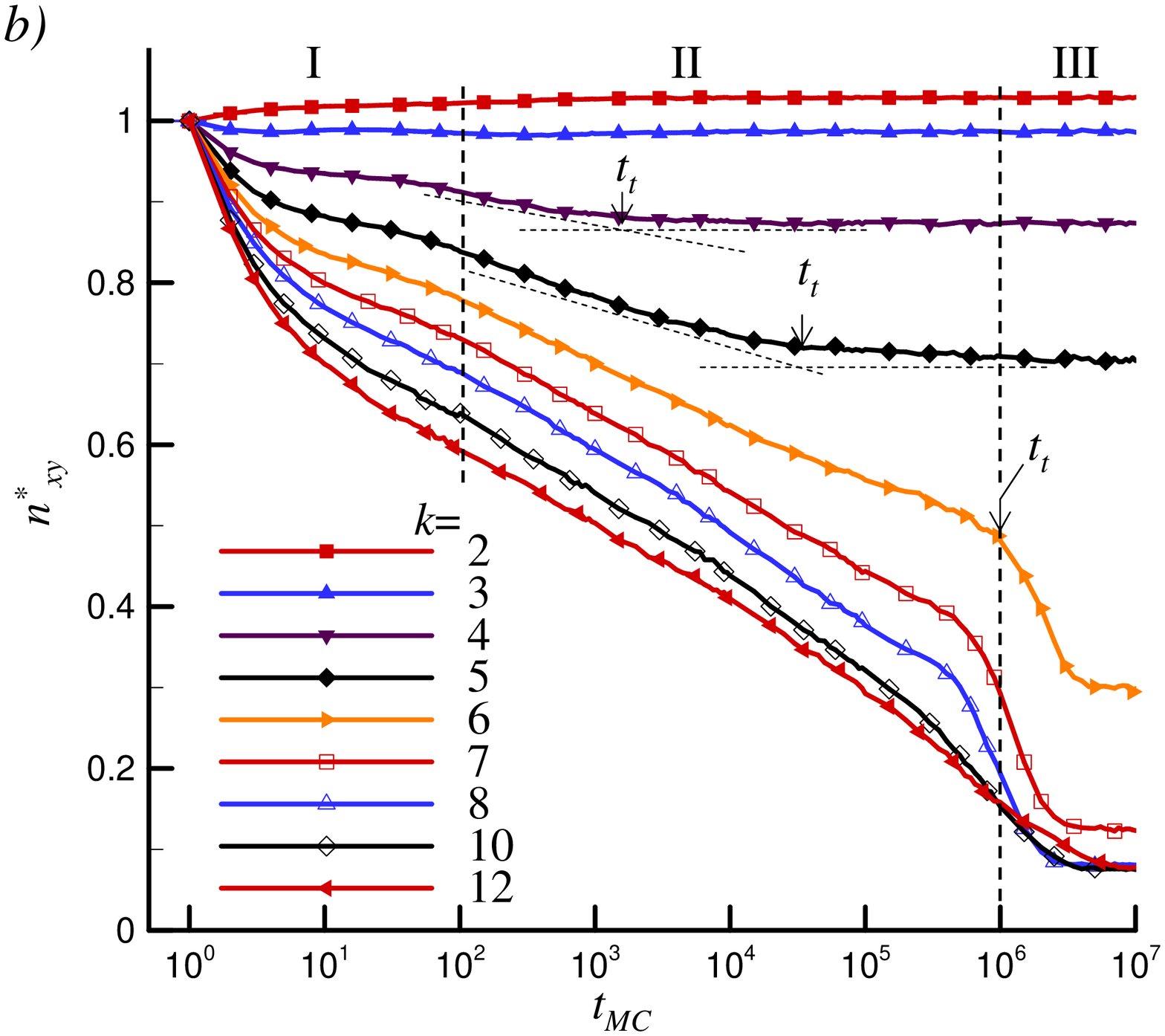}
\caption{Normalized number of (a) intraspecific, $n^*=n/n^i$, and (b)  interspecific, $n^*_{xy}=n_{xy}/n^i_{xy}$, contacts versus the time $t_{MC}$ for different values of length of the $k$-mers. Here, $n^i$ and $n^i_{xy}$ are the corresponding initial values at $t_{MC}=1$, while $L=256$ and the initial r configuration were used. The concentration of $k$-mers corresponds to the jamming state~\cite{Lebovka2017}. The values of (a)  $n^*_f$ and (b) $t_t$ correspond to the final values at $t_{MC}=10^7$ and to the transition period, respectively.
\label{fig:f05ab}}
\end{figure}

Figure~\ref{fig:f05ab} compares the kinetics of the changes in the number of (a) intraspecific and  (b) interspecific contacts for different values of $k$. Here, for the convenience of presentation, the data have been normalized to their initial values at $t_{MC}=1$.
%%%%%%%%%%%%%%%%%%%%%%%%%%%%%%%%%%%%%%%%%%%%%%%%%%%%%%%%%%%%%%%%%%%%%%%%%%%%%%%%%%%%%%%%%%%%%%%%%%%%%%%%%%%%%%%%%%%%%%%%%%%%%%%%%%%%%%%%%%%%%%%%%%%
%%%%%%%%%%%%%%%%%%%%%%%%%%%%%%%%%%%%%%%%%%%%%%%%%%%%%%%%%%%%%%%%%%%%%%%%%%%%%%%%%%%%%%%%%%%%%%%%%%%%%%%%%%%%%%%%%%%%%%%%%%%%%%%%%%%%%%%%%%%%%%%%%%%

The normalized number of intraspecific contacts, $n^*$, grows during the time period $t_{MC}=1-10$ for all values of $k$ (Fig.~\ref{fig:f05ab}a). During the intermediate stage II and the final stage III, the time dependencies were different for different values of $k$. The most remarkable feature was the presence of steps at the boundary between stages II and III for $k=6,7,$ and $8$ and their absence for other values of $k$. The dependence of the normalized number of intraspecific contacts, $n^*_f$, versus $k$ for relatively long kinetic MC simulations, $t_{MC}=10^7$, is presented in Fig.~\ref{fig:f06_contacts}. The maximum value of $n^*_f$ was observed at $k=6$.

\begin{figure}[htbp]
  \centering
  \includegraphics[width=0.9\columnwidth]{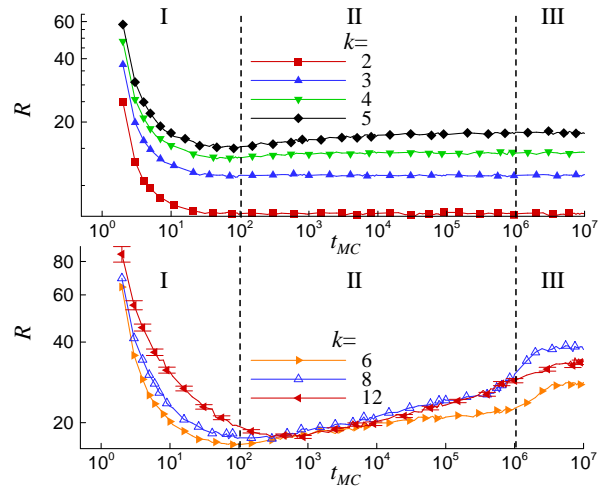}
\caption{Shift ratio, $R$, versus the time $t_{MC}$ for different values of length of the $k$-mers. Here, $L=256$ and the initial r configuration were used. The concentration of $k$-mers corresponds to the jamming state~\cite{Lebovka2017}.
\label{fig:f07_R}}
\end{figure}

\begin{figure}[htbp]
  \centering
  \includegraphics[width=0.9\columnwidth]{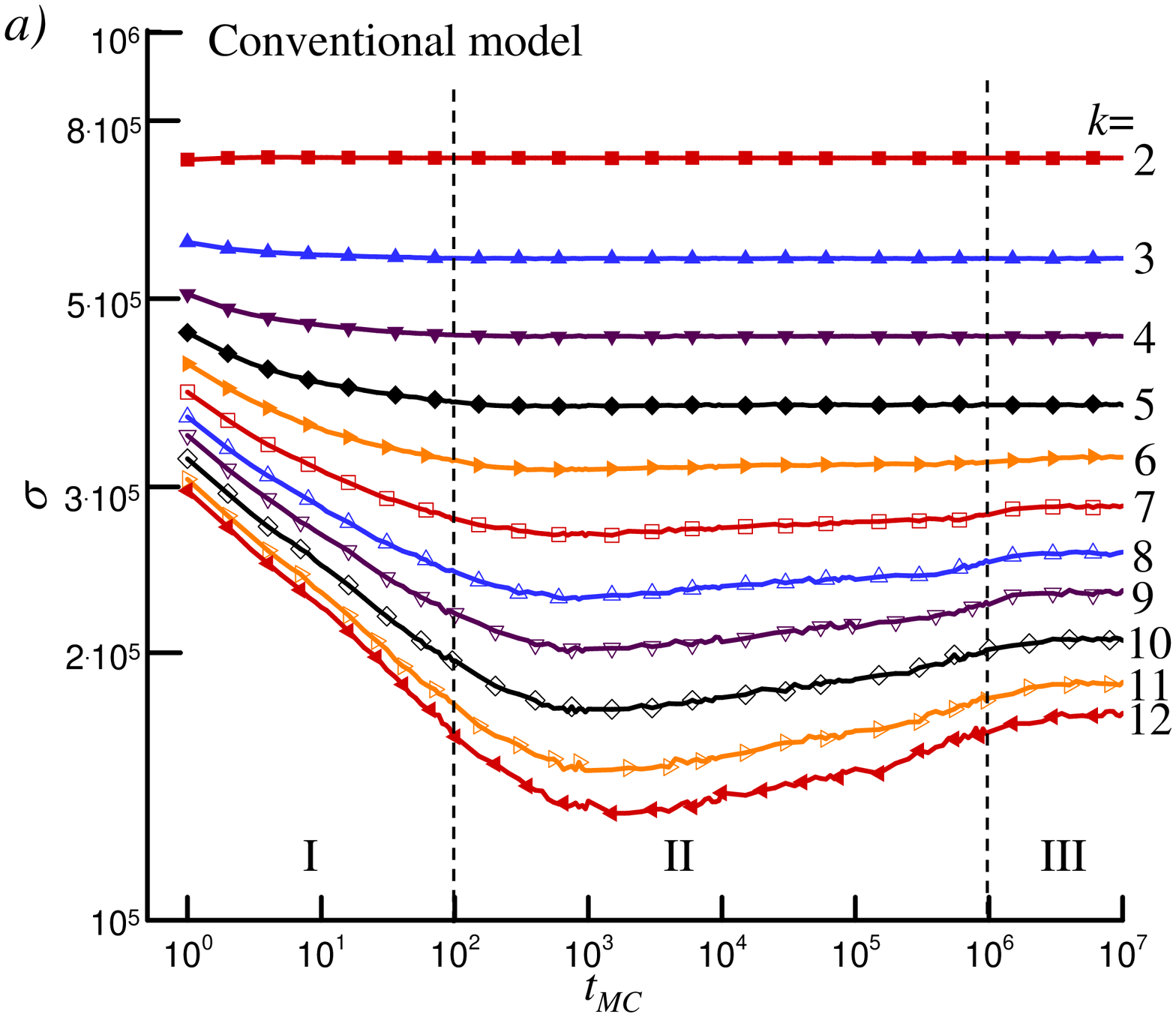}
    \includegraphics[width=0.9\columnwidth]{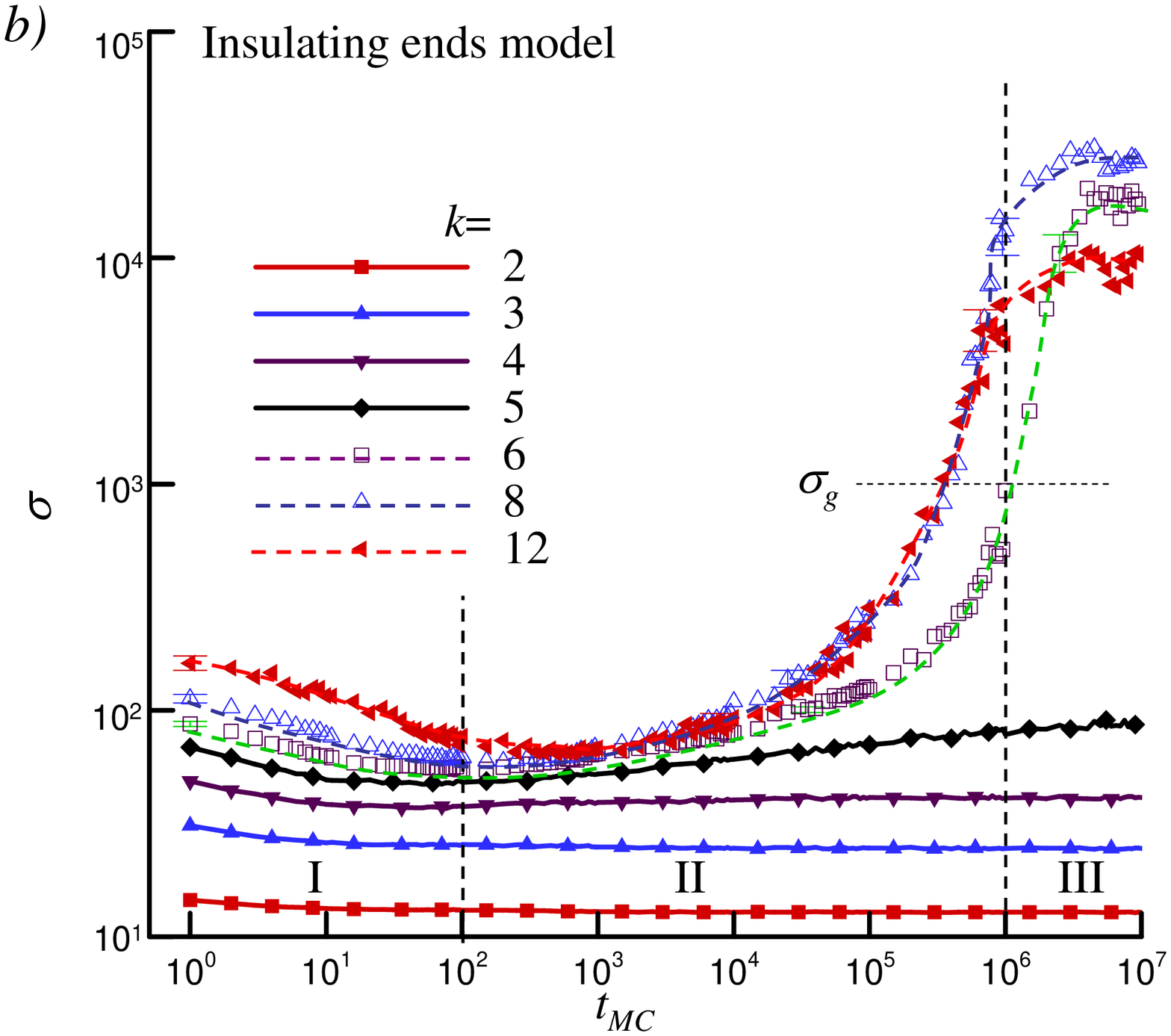}
\caption{Electrical conductivity, $\sigma$, versus the time $t_{MC}$ for different values of length of the $k$-mers for (a) conventional and (b) insulating ends models. Here, $L=256$ and the initial r configuration were used. The concentration of $k$-mers corresponds to the jamming state~\cite{Lebovka2017}.
\label{fig:f08_cond}}
\end{figure}

The kinetics of the changes in the normalized number of interspecific contacts, $n_{xy}^*$, were also different for various values of $k$  (Fig.~\ref{fig:f05ab}b). These changes were insignificant for $k=2,3$. For other values of $k$, the values of $n_{xy}^*$ decreased during the initial period of time and then stabilized after a transition period, $t_t$. The transition periods were $t_t\approx 2\times 10^3$ for $k=4$, $t_t\approx 3\times 10^4$ for $k=5$, and $t_t\approx 3\times 10^6$ for $k\geq6$.
The kinetics of changes in the shift ratios $R(t_{MC})$ and electrical conductivity $\sigma(t_{MC})$ for different values of $k$ are presented in Fig.~\ref{fig:f07_R} and Fig.~\ref{fig:f08_cond}, respectively.
%%%%%%%%%%%%%%%%%%%%%%%%%%%%%%%%%%%%%%%%%%%%%%%%%%%%%%%%%%%%%%%%%%%%%%%%%%%%%%%%%%%%%%%%%%%%%%%%%%%%%%%%%%%%%%%%%%%%%%%%%%%%%%%%%%%%%%%%%%%%%%%%%%%
%%%%%%%%%%%%%%%%%%%%%%%%%%%%%%%%%%%%%%%%%%%%%%%%%%%%%%%%%%%%%%%%%%%%%%%%%%%%%%%%%%%%%%%%%%%%%%%%%%%%%%%%%%%%%%%%%%%%%%%%%%%%%%%%%%%%%%%%%%%%%%%%%%%
For all values of $k$ the  values of $R$ decreased significantly during the initial stage I (at $t_{MC} <10^2$) (Fig.~\ref{fig:f07_R}). In the initial jammed state,  shifts along the transverse axes of the $k$-mers are not possible (i.e., $R=\infty$) and the initial decrease of $R$ evidently reflects a rapid transition of the system into a non-jammed state. For $k\leq 5$, the values of $R$ stabilized at some level within the intermediate (II) and final (III) stages. For $k\geq 6$, distinct minima within stage II and the steps at the boundary between stages II and III were observed in the $R(t_{MC})$ dependencies. The observed steps correlate with the similar steps observed in the $n_{xy}^*(t_{MC})$ and $n_{xy}^*(t_{MC})$ dependencies (Fig.~\ref{fig:f05ab}). These steps can be explained by the starting of the process of stripe domain formation. It is evident, that the formation of domains of similarly oriented $k$-mers accompanies the restriction of the shifts along the transverse axis.

The kinetics of the changes in the electrical conductivity, $\sigma(t_{MC})$ were significantly different for the conventional (Fig.~\ref{fig:f08_cond}a) and insulating ends (Fig.~\ref{fig:f08_cond}b) models. Note that, for isotropic deposition, the mean geometric conductivity, $\sigma_g=\sqrt{\sigma_m \sigma_p}=10^3$ approximately corresponds to the value of the electrical conductivity at the percolation threshold \cite{Tarasevich2016}.
%%%%%%%%%%%%%%%%%%%%%%%%%%%%%%%%%%%%%%%%%%%%%%%%%%%%%%%%%%%%%%%%%%%%%%%%%%%%%%%%%%%%%%%%%%%%%%%%%%%%%%%%%%%%%%%%%%%%%%%%%%%%%%%%%%%%%%%%%%%%%%%%%%%
%%%%%%%%%%%%%%%%%%%%%%%%%%%%%%%%%%%%%%%%%%%%%%%%%%%%%%%%%%%%%%%%%%%%%%%%%%%%%%%%%%%%%%%%%%%%%%%%%%%%%%%%%%%%%%%%%%%%%%%%%%%%%%%%%%%%%%%%%%%%%%%%%%%

For the conventional model, the electrical conductivity, $\sigma$, during the diffusion evolution of the system, was noticeably greater compared with $\sigma_g$ for all values of $k$ (Fig.~\ref{fig:f08_cond}a). This reflected the presence of percolation during the evolution. At the initial moment (in the jamming state), the percolation concentration, $p_c$, is always smaller than the jamming concentration, $p_j$, for the studied range $k \in [ 2, 12 ]$ \cite{Tarasevich2012PRE}. The temporal changes of electrical conductivity were insignificant for $k<5$. For $k\geq6$ the  values of $\sigma$ decreased noticeably during the initial stage I (for $t <100$) and distinct minima at $t\approx 10^3$ were observed.
For the insulating ends (Fig.~\ref{fig:f08_cond}b) models, significant changes of electrical conductivity, $\sigma$, with a sharp transition above $\sigma_g$ were only observed at the boundary between stages II and III for relatively long $k$-mers ($k\geq 6$).

\begin{figure}[htbp]
  \centering
    \includegraphics[width=0.9\columnwidth]{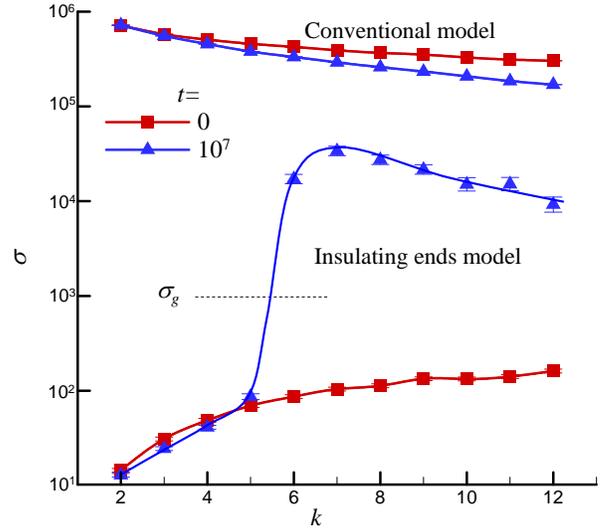}
\caption{Electrical conductivity, $\sigma$, versus the length of $k$-mers for different values of $t_{MC}$ for conventional and insulating ends models. Here, $L=256$ and the initial r configuration were used. The concentration of $k$-mers corresponds to the jamming state~\cite{Lebovka2017}.
\label{fig:f09_sigma}}
\end{figure}

Figure~\ref{fig:f09_sigma} compares the electrical conductivity, $\sigma$, versus the $k$ dependencies at the initial jamming state ($t_{MC}=1$) and at the final state, after a relatively long time ($t_{MC}=10^7$) for the conventional and insulating ends models. On the one hand, for the conventional model, the value of $\sigma$ decreased with increase of $k$, in both the initial and final states. This  naturally reflected the decrease in the value of the jamming concentration for the longer $k$-mers \cite{Lebovka2011}. The diffusion evolution also resulted in a decrease of $\sigma$. On the other hand, for the insulating ends model, during the initial stage, the value of $\sigma$ decreased with increase of $k$  but was always below the level of the mean geometric conductivity, $\sigma_g$. We can conclude that, during the initial stage, a suppression of the percolation for the insulating ends model is caused by the  presence of insulating ends. So, the presence of the insulating ends has a significant impact on the connectivity between the x and y stacks. During the diffusion, the formation of the stripes for $k\geq 6$ resulted in a restoration of percolation through the system via the connectivity inside the large domains of  similarly oriented $k$-mers.
%%%%%%%%%%%%%%%%%%%%%%%%%%%%%%%%%%%%%%%%%%%%%%%%%%%%%%%%%%%%%%%%%%%%%%%%%%%%%%%%%%%%%%%%%%%%%%%%%%%%%%%%%%%%%%%%%%%%%%%%%%%%%%%%%%%%%%%%%%%%%%%%%%%
%%%%%%%%%%%%%%%%%%%%%%%%%%%%%%%%%%%%%%%%%%%%%%%%%%%%%%%%%%%%%%%%%%%%%%%%%%%%%%%%%%%%%%%%%%%%%%%%%%%%%%%%%%%%%%%%%%%%%%%%%%%%%%%%%%%%%%%%%%%%%%%%%%%

\subsection{Effects of the size of the system:  \texorpdfstring{$L=128-2048$}{L=128-2048} \label{subsec:L2048}}
\begin{figure}[htbp]
  \centering
    \includegraphics[width=\linewidth]{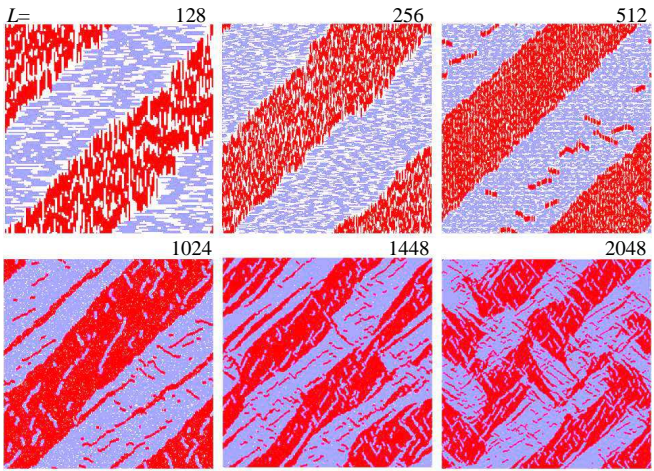}
\caption{Examples of patterns at $t_{MC}=10^7$  for $k=12$ and different sizes of the lattice $L=128-2048$. Here the initial r configuration was used. The concentration of $k$-mers corresponds to the jamming state. \label{fig:f10_k12_Patterns_Scaling}}
\end{figure}

Figure~\ref{fig:f10_k12_Patterns_Scaling} presents examples of the patterns formed with a fixed length of $k$-mers, $k=12$ and different sizes of the system, $L=128-2048$ at $t_{MC}=10^7$. The initial r configuration was used. The fairly ideal stripe domains were only observed for $L=128$ and $L=256$.
For larger lattices starting from $L=512$,  stacks consisting of $k$-mers of perpendicular orientation appeared inside the stripe domains and their concentration increased with increase of $L$. We believe that these alien stacks may be found to disappear when $t_{MC} \to \infty$. Unfortunately, we are not in a position to perform any direct verification  of this suggestion because the simulation is very time-consuming.
%%%%%%%%%%%%%%%%%%%%%%%%%%%%%%%%%%%%%%%%%%%%%%%%%%%%%%%%%%%%%%%%%%%%%%%%%%%%%%%%%%%%%%%%%%%%%%%%%%%%%%%%%%%%%%%%%%%%%%%%%%%%%%%%%%%%%%%%%%%%%%%%%%%
%%%%%%%%%%%%%%%%%%%%%%%%%%%%%%%%%%%%%%%%%%%%%%%%%%%%%%%%%%%%%%%%%%%%%%%%%%%%%%%%%%%%%%%%%%%%%%%%%%%%%%%%%%%%%%%%%%%%%%%%%%%%%%%%%%%%%%%%%%%%%%%%%%%

The example of the evolution over time of the patterns for $k=12$ and $L=2048$ is presented in Video~\ref{vid:f11_k12_Patterns_2048}.
In this case, an evident coarsening of clusters elongated along the diagonals of the lattice was observed at $t_{MC}=10^5$--$10^6$.
%%%%%%%%%%%%%%%%%%%%%%%%%%%%%%%%%%%%%%%%%%%%%%%%%%%%%%%%%%%%%%%%%%%%%%%%%%%%%%%%%%%%%%%%%%%%%%%%%%%%%%%%%%%%%%%%%%%%%%%%%%%%%%%%%%%%%%%%%%%%%%%%%%%
\begin{video}[htbp]
  \centering
  \includegraphics[width=\linewidth]{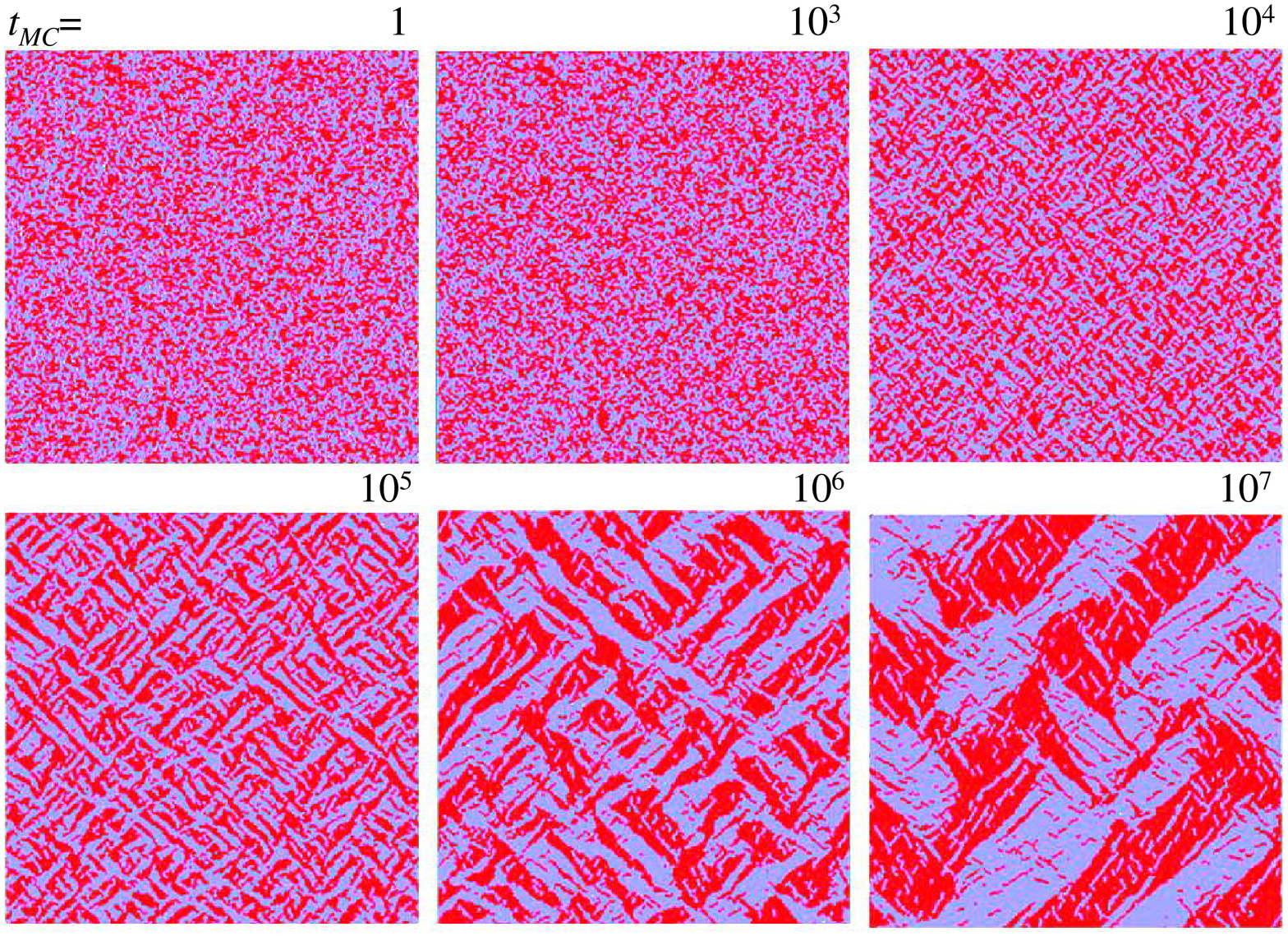}
  \setfloatlink{videok12L2048.avi}
\caption{Examples of patterns  at different moments in the time, $t_{MC}$. Here, $L=2048$, $k=12$ and the initial r configuration were used. The concentration of $k$-mers corresponds to the jamming state. \label{vid:f11_k12_Patterns_2048}}
\end{video}
%%%%%%%%%%%%%%%%%%%%%%%%%%%%%%%%%%%%%%%%%%%%%%%%%%%%%%%%%%%%%%%%%%%%%%%%%%%%%%%%%%%%%%%%%%%%%%%%%%%%%%%%%%%%%%%%%%%%%%%%%%%%%%%%%%%%%%%%%%%%%%%%%%%

\begin{figure}[htbp]
  \centering
  \includegraphics[width=0.9\columnwidth]{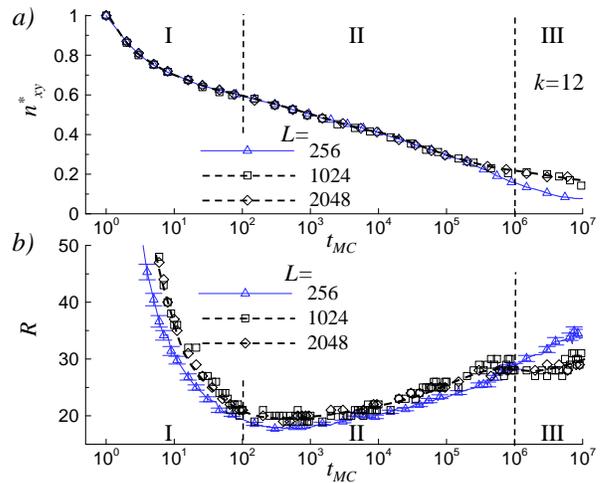}
\caption{Normalized number of (a) interspecific contacts, $n^*_{xy}=n_{xy}/n^i_{xy}$, and (b) the shift ratio, $R$, versus the time $t_{MC}$ for $k=12$ and different sizes of the system, $L$. Here, $n^i_{xy}$ is the initial value at $t_{MC}=1$ and the initial r configuration were used. The concentration of $k$-mers corresponds to the jamming state.
\label{fig:f12_k12ab_Scaling}}
\end{figure}

The defects  inside the stripe domains were strongly stabilized and did not disappear during diffusion evolution. Figure~\ref{fig:f12_k12ab_Scaling} compares the kinetics of the changes in (a) the normalized number of interspecific contacts, $n^*_{xy}=n_{xy}/n^i_{xy}$, and (b) the shift ratio, $R$, for $k=12$ and for different sizes of the system, $L$. The data evidenced the presence of fairly insignificant scaling within the studied range of $L$. Therefore, we can assume that the results for $L=256$ presented earlier are qualitatively correct for systems of larger size.
%%%%%%%%%%%%%%%%%%%%%%%%%%%%%%%%%%%%%%%%%%%%%%%%%%%%%%%%%%%%%%%%%%%%%%%%%%%%%%%%%%%%%%%%%%%%%%%%%%%%%%%%%%%%%%%%%%%%%%%%%%%%%%%%%%%%%%%%%%%%%%%%%%%
%%%%%%%%%%%%%%%%%%%%%%%%%%%%%%%%%%%%%%%%%%%%%%%%%%%%%%%%%%%%%%%%%%%%%%%%%%%%%%%%%%%%%%%%%%%%%%%%%%%%%%%%%%%%%%%%%%%%%%%%%%%%%%%%%%%%%%%%%%%%%%%%%%%

\subsection{Comparison of the different initial configurations\label{subsec:comparisionIC}}

In order to clarify the mechanism of formation of the diagonal stripe domains for $k\geq 6$, the effects of different initial configurations on such pattern formation were also studied. A strong impact of the initial configuration on the evolution of patterns and the kinetics of the main parameters of the system was observed. Video~\ref{vid:f13_k12_Patterns_vd} compares the time evolution of the patterns for initial stripe configurations: v configuration (upper row) and d configuration (bottom row). Here, $L=256$, $k=12$ and the concentration of particles corresponds to the jamming concentration for the initial r  configuration. It is clearly evident, that the initial unstable v configuration gradually transforms to diagonal stripe patterns, whereas the initial stable d configuration remains almost unchanged. We can conclude that the pattern of diagonal stripe domains is an attractor, i.e., any spatial distribution of $k$-mers tends to transform into diagonal stripes when $k\geq 6$.
%%%%%%%%%%%%%%%%%%%%%%%%%%%%%%%%%%%%%%%%%%%%%%%%%%%%%%%%%%%%%%%%%%%%%%%%%%%%%%%%%%%%%%%%%%%%%%%%%%%%%%%%%%%%%%%%%%%%%%%%%%%%%%%%%%%%%%%%%%%%%%%%%%%
\begin{video}[htbp]
  \centering
  \includegraphics[width=\linewidth]{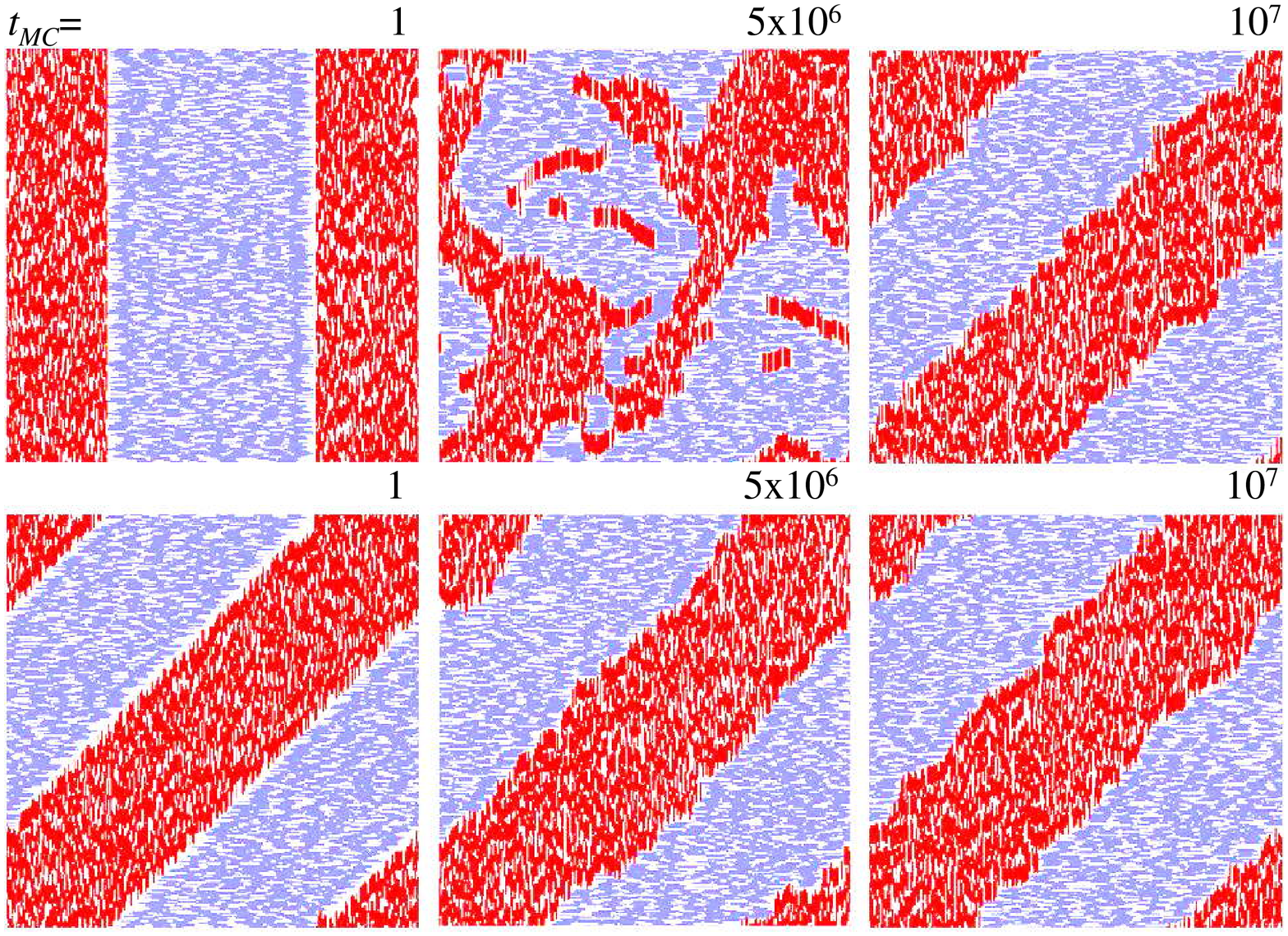}
  \setfloatlink{videokk12vd.avi}
\caption{Examples of patterns  at different moments in the time, $t_{MC}$. Here, $L=256$, $k=12$, initial v configuration (upper row) and d configuration (bottom row) were used. Concentration of $k$-mers corresponds to the jamming state of r configuration~\cite{Lebovka2017}.
\label{vid:f13_k12_Patterns_vd}}
\end{video}
%%%%%%%%%%%%%%%%%%%%%%%%%%%%%%%%%%%%%%%%%%%%%%%%%%%%%%%%%%%%%%%%%%%%%%%%%%%%%%%%%%%%%%%%%%%%%%%%%%%%%%%%%%%%%%%%%%%%%%%%%%%%%%%%%%%%%%%%%%%%%%%%%%%

\begin{figure}[htbp]
  \centering
   \includegraphics[width=0.9\columnwidth]{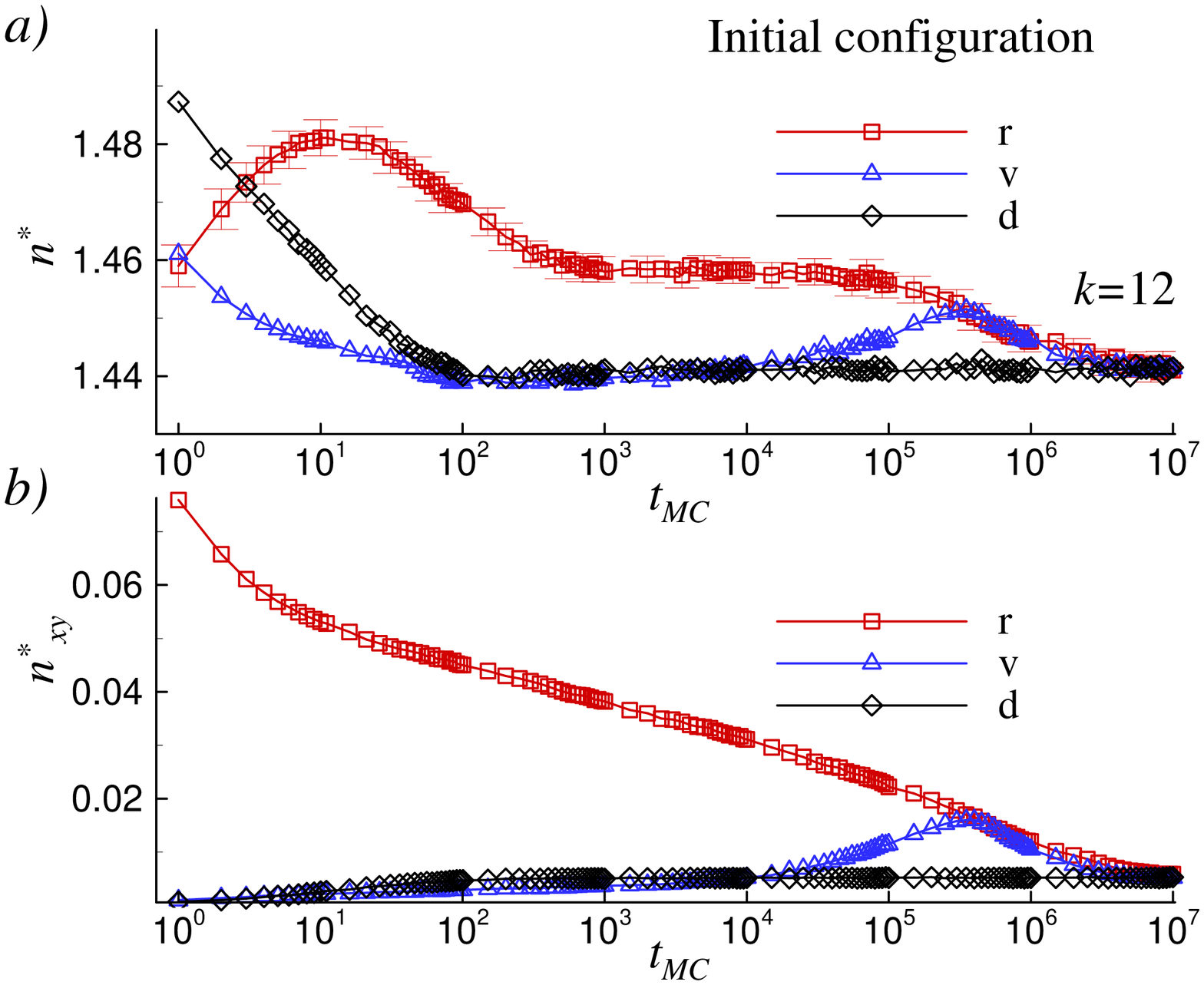}
  \includegraphics[width=0.9\columnwidth]{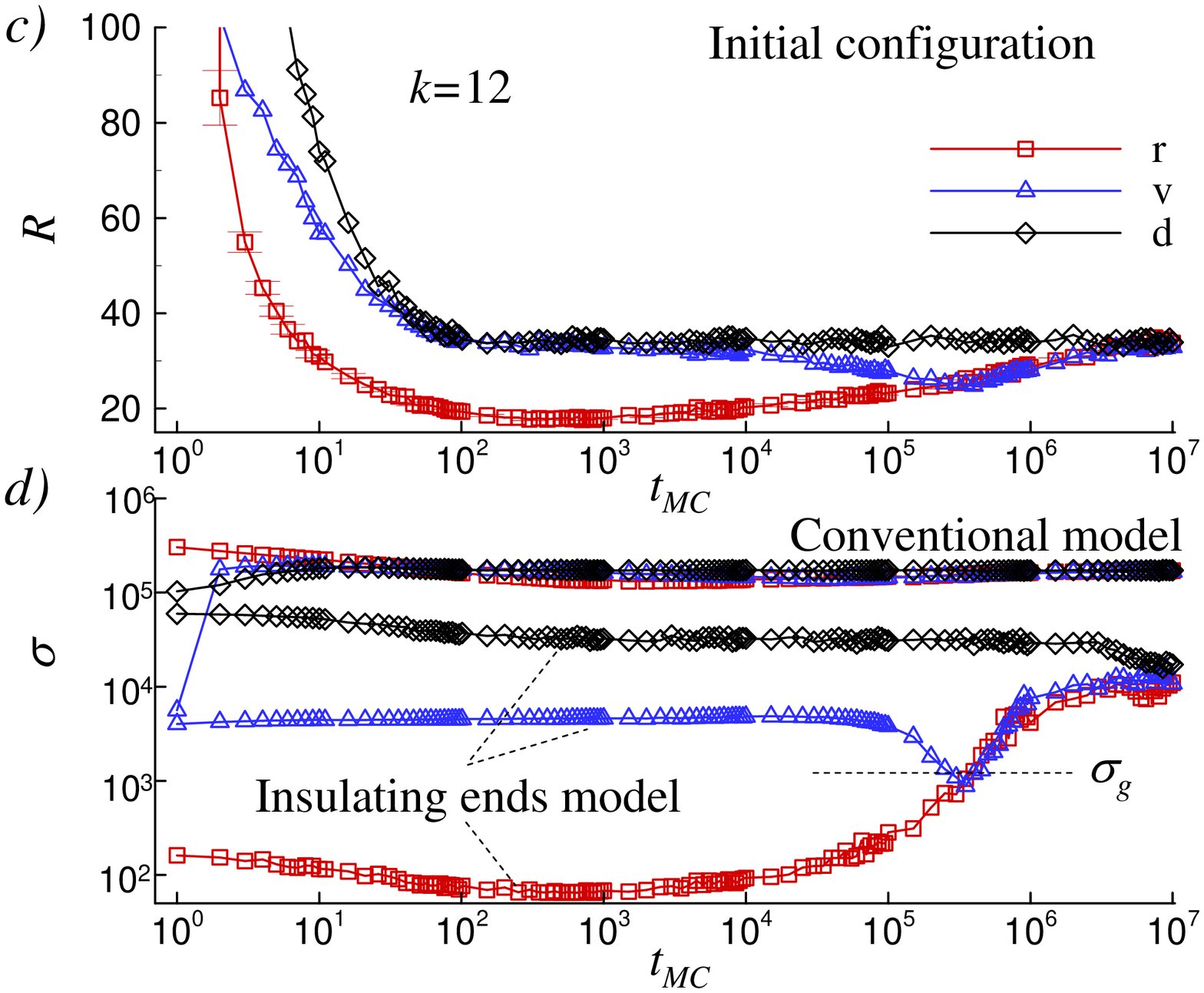}
\caption{Normalized number of (a) intraspecific contacts, $n^*=n/n^i$, (b) interspecific contacts, $n^*_{xy}=n_{xy}/n^i_{xy}$, (c) shift ratio, $R$, and (d) electrical conductivity, $\sigma$, versus time $t_{MC}$ for $k=12$. Here, $n^i$ and $n^i_{xy}$ are the corresponding initial values at $t_{MC}=1$, while $L=256$ and different initial r,v, and d configurations were used. The concentration of $k$-mers corresponds to the jamming state of the r configuration~\cite{Lebovka2017}.\label{fig:f14abcd}}
\end{figure}

Figure~\ref{fig:f14abcd} compares the kinetics of changes in (a) the  normalized number of intraspecific contacts, $n^*$, (b) the interspecific contacts, $n^*_{xy}$, (c) the shift ratio, $R$,  and (d) the electrical conductivity, $\sigma$, for $k=12$ and $L=256$. It is remarkable that,
for various initial configurations, the initial values and the evolution of all parameters were fairly  different. %%%%%%%%%%%%%%%%%%%%%%%%%%%%%%%%%%%%%%%%%%%%%%%%%%%%%%%%%%%%%%%%%%%%%%%%%%%%%%%%%%%%%%%%%%%%%%%%%%%%%%%%%%%%%%%%%%%%%%%%%%%%%%%%%%%%%%%%%%%%%%%%%%%
%%%%%%%%%%%%%%%%%%%%%%%%%%%%%%%%%%%%%%%%%%%%%%%%%%%%%%%%%%%%%%%%%%%%%%%%%%%%%%%%%%%%%%%%%%%%%%%%%%%%%%%%%%%%%%%%%%%%%%%%%%%%%%%%%%%%%%%%%%%%%%%%%%%

Their final values were almost the same irrespective of the initial configuration. Note that, for the initial v and d configuration, for the insulation ends model, the electrical conductivity was above the mean geometric conductivity, $\sigma_g$. This reflected the presence of percolation along the stripe domains. It is also interesting that all the studied parameters illustrated extrema at $t_{MC}\approx 2\times10^5$, i.e., just near the boundary between stages II and III. This corresponds to the critical transformation of the vertical stripe into the diagonal stripe (see pattern for initial v configuration at $t_{MC}=5\times10^5$, Fig.~\ref{fig:f12_k12ab_Scaling}).
%%%%%%%%%%%%%%%%%%%%%%%%%%%%%%%%%%%%%%%%%%%%%%%%%%%%%%%%%%%%%%%%%%%%%%%%%%%%%%%%%%%%%%%%%%%%%%%%%%%%%%%%%%%%%%%%%%%%%%%%%%%%%%%%%%%%%%%%%%%%%%%%%%%
\begin{figure}[htbp]
  \centering
    \includegraphics[width=\linewidth]{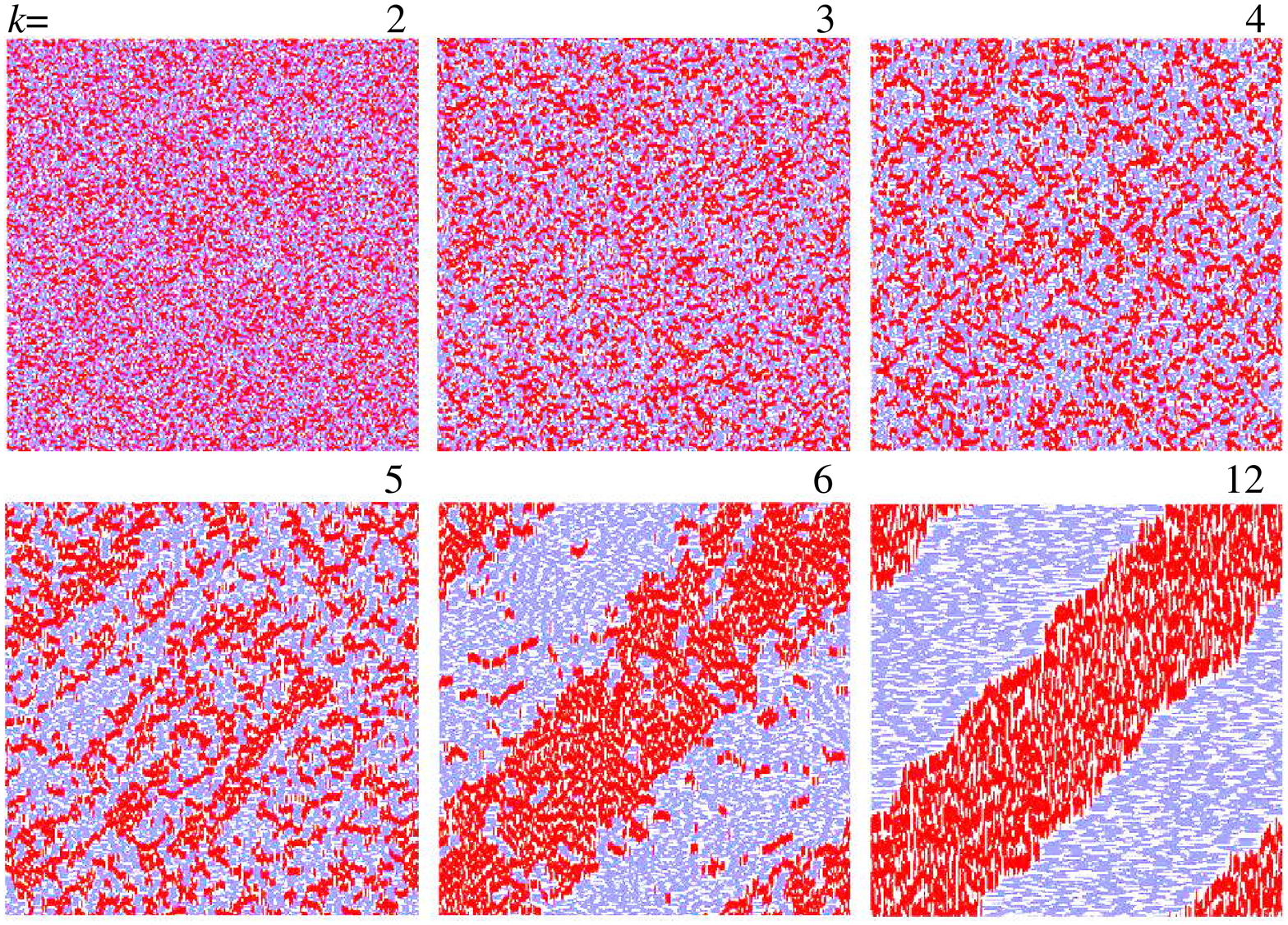}
\caption{Examples of patterns at $t_{MC}=10^7$  for  different values of length of the $k$-mers.
Here, $L=256$ and the initial d configuration were used. The concentration of $k$-mers corresponds to the jamming state for the initial r configuration~\cite{Lebovka2017}. \label{f15_Patterns_Final_D_all}}
\end{figure}
%%%%%%%%%%%%%%%%%%%%%%%%%%%%%%%%%%%%%%%%%%%%%%%%%%%%%%%%%%%%%%%%%%%%%%%%%%%%%%%%%%%%%%%%%%%%%%%%%%%%%%%%%%%%%%%%%%%%%%%%%%%%%%%%%%%%%%%%%%%%%%%%%%%

%%%%%%%%%%%%%%%%%%%%%%%%%%%%%%%%%%%%%%%%%%%%%%%%%%%%%%%%%%%%%%%%%%%%%%%%%%%%%%%%%%%%%%%%%%%%%%%%%%%%%%%%%%%%%%%%%%%%%%%%%%%%%%%%%%%%%%%%%%%%%%%%%%%
\begin{figure}[htbp]
  \centering
  \includegraphics[width=0.9\columnwidth]{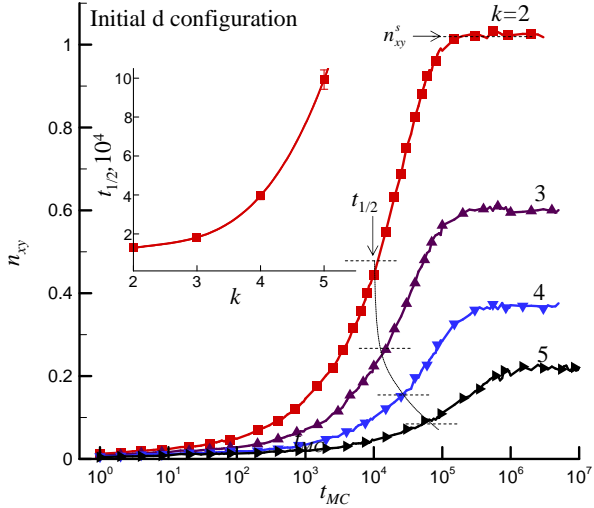}
\caption{Number of interspecific, $n_{xy}$, contacts versus time $t_{MC}$ for different values of $k$. Here, $L=256$, and the initial d configuration were used. The concentration of $k$-mers corresponds to the jamming state for the initial r configuration~\cite{Lebovka2017}. Inset shows the half-time $t_{1/2}$ required to attain the level $0.5n_{xy}^s$.
\label{fig:f16_n_vs_t_D_all}}
\end{figure}
%%%%%%%%%%%%%%%%%%%%%%%%%%%%%%%%%%%%%%%%%%%%%%%%%%%%%%%%%%%%%%%%%%%%%%%%%%%%%%%%%%%%%%%%%%%%%%%%%%%%%%%%%%%%%%%%%%%%%%%%%%%%%%%%%%%%%%%%%%%%%%%%%%%

%%%%%%%%%%%%%%%%%%%%%%%%%%%%%%%%%%%%%%%%%%%%%%%%%%%%%%%%%%%%%%%%%%%%%%%%%%%%%%%%%%%%%%%%%%%%%%%%%%%%%%%%%%%%%%%%%%%%%%%%%%%%%%%%%%%%%%%%%%%%%%%%%%%
\begin{figure}[htbp]
  \centering
  \includegraphics[width=0.9\columnwidth]{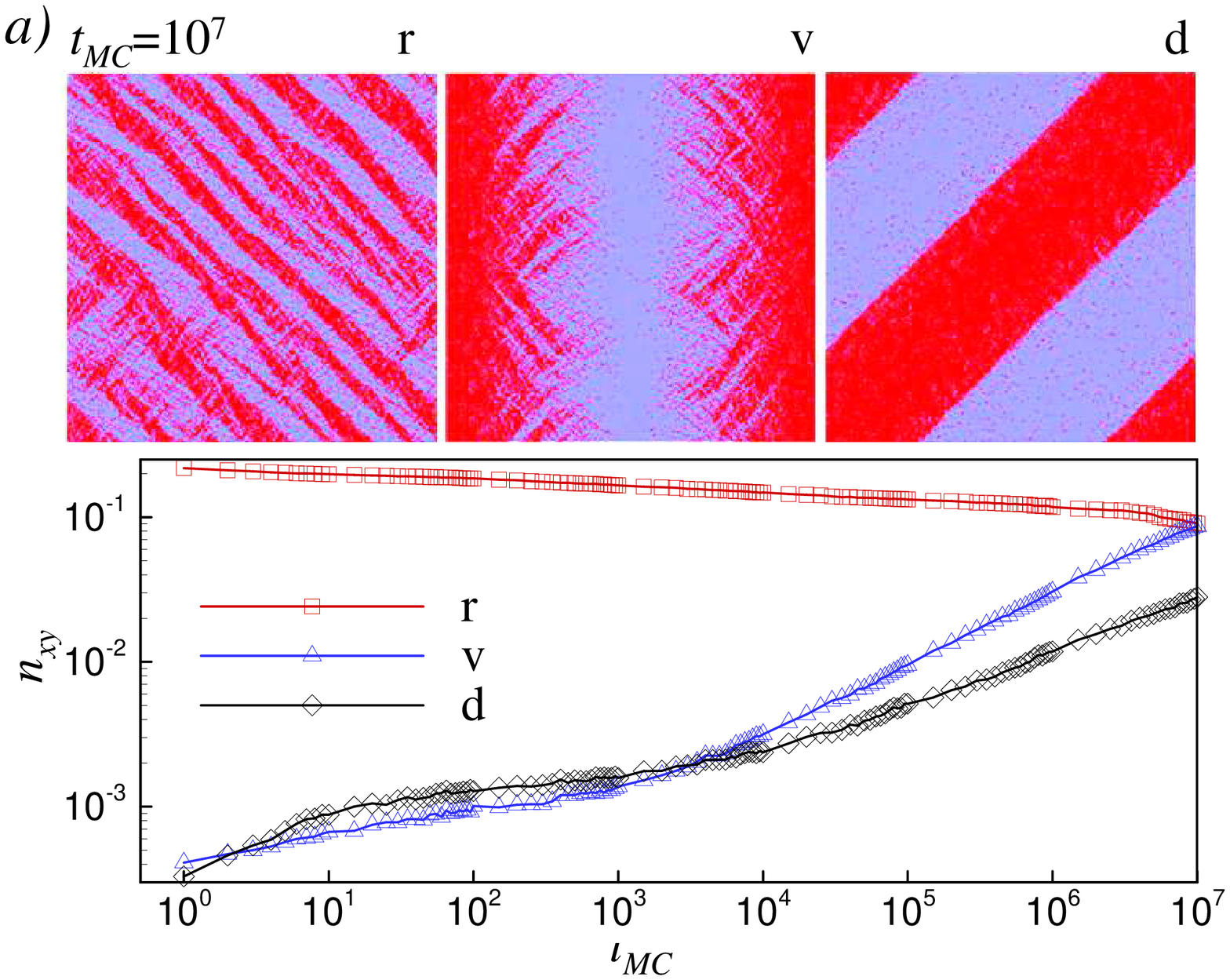}
    \includegraphics[width=0.9\columnwidth]{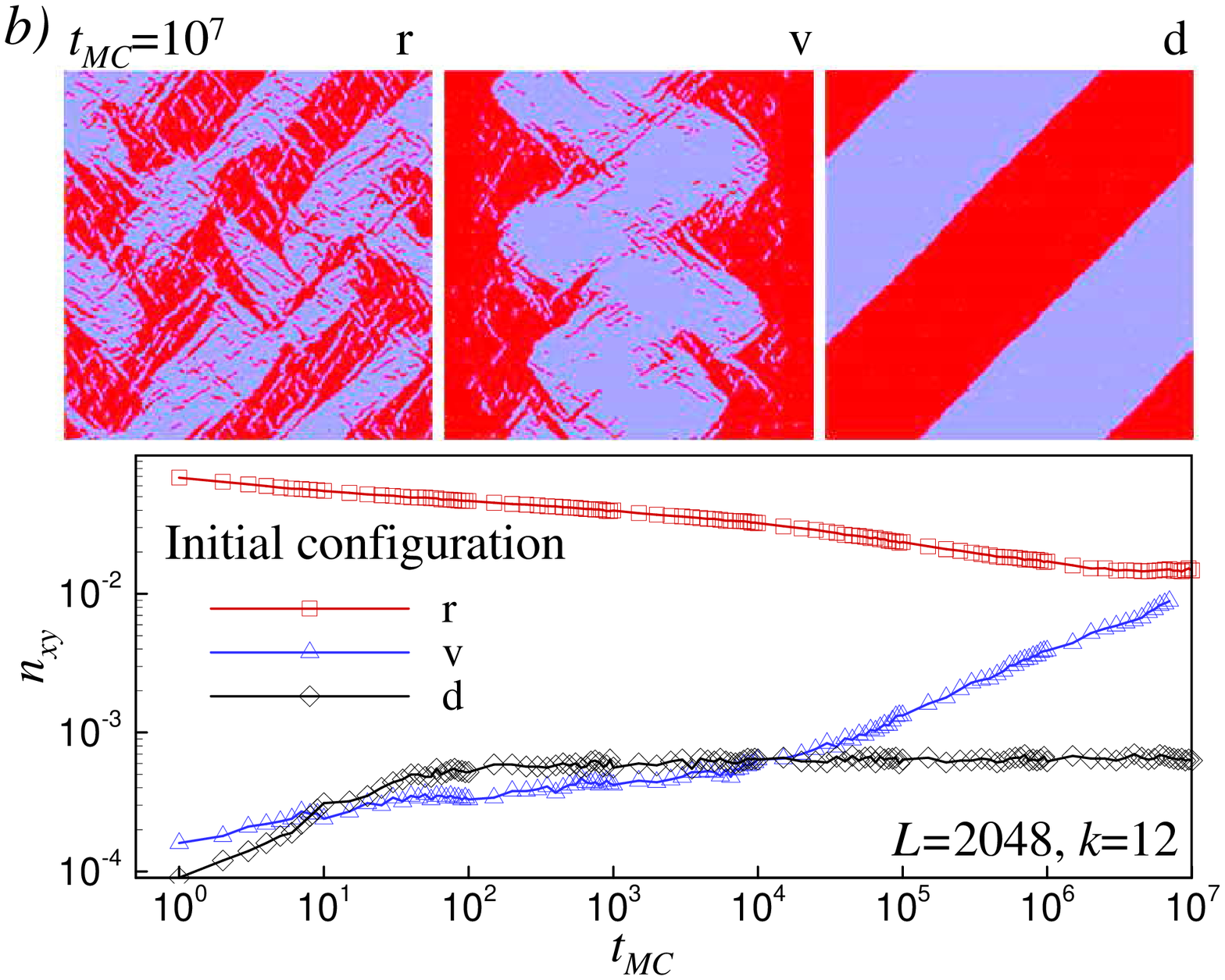}
\caption{Number of interspecific, $n_{xy}$, contacts versus time $t_{MC}$ for initial r, v and d configurations, $L=2048$, for (a) $k=6$ and (b) $k=12$. The concentration of $k$-mers corresponds to the jamming state for the initial r configuration~\cite{Lebovka2017}. The corresponding patterns at $t_{MC}=10^7$ are also shown.
\label{fig:f17}}
\end{figure}
%%%%%%%%%%%%%%%%%%%%%%%%%%%%%%%%%%%%%%%%%%%%%%%%%%%%%%%%%%%%%%%%%%%%%%%%%%%%%%%%%%%%%%%%%%%%%%%%%%%%%%%%%%%%%%%%%%%%%%%%%%%%%%%%%%%%%%%%%%%%%%%%%%%

Similar effects were observed for other values of $k$ in the range of $6-12$. At smaller values of $k$, the choice of initial d configuration did not allow the diagonal stripe patterns to stabilize. Figure~\ref{f15_Patterns_Final_D_all} presents examples of such patterns at $t_{MC}=10^7$  for  different values of $k$. It is interesting to note that for the initial d configuration the time evolution of the normalized number of interspecific contacts, $n^*_{xy}$, was insignificant for long $k$-mers ($k=6-12$) (e.g., see Fig.~\ref{fig:f14abcd}b).  Nevertheless, it was fairly significant for shorter $k$-mers Fig.~\ref{fig:f16_n_vs_t_D_all}.

The value of $n^*_{xy}$ increased during the diffusion evolution of the system, and saturated for a long time at the level of $n_{xy}^s$.
The inset in Fig.~\ref{fig:f16_n_vs_t_D_all} shows the half-time $t_{1/2}$ required to attain $0.5n_{xy}^s$. The most rapid kinetics were observed for small $k$-mer lengths.

Finally, figure~\ref{fig:f17} compares the kinetics of changes in the normalized number of intraspecific contacts, $n^*$,
for initial r, v and d configurations for large systems, $L=2048$, (a) $k=6$ and (b) $k=12$.
For $k=6$, the systems with initial v and d configurations progressively relaxed to the state of a system with an initial r configuration with numerous defected diagonal stripes (Fig.~\ref{fig:f17}a).

For $k=12$, the gradual  relaxation to the defected diagonal stripes was only observed for the initial v configuration (Fig.~\ref{fig:f17}b). The initial d configuration was stabilized and only point defects were observed after long MC simulations. So, for large systems such as, $L=2048$, different non-equilibrium steady states can be realized, with a dependence on the initial configurations and the value of~$k$.

\section{Conclusion\label{sec:conclusion}}
The diffusion-driven self-assembly of rod-like particles ($k$-mers) oriented along $x$ and $y$ directions on a square lattice was studied by means of MC simulation. Isotropic orientations for the $k$-mers was assumed at the starting point. We considered only very dense systems, i.e., systems at the jamming concentration. The initial jamming state was produced by RSA, then, the $k$-mers were allowed to diffuse. The typical time of deposition was supposed to be much less than the typical time of relaxation.  In such concentrated systems, only the translational diffusion of particles is possible, whereas rotational diffusion is completely inhibited. The length of the $k$-mers (and therefore their, aspect ratio)  was varied from $2$ to $12$. The size of lattice, $L$, was varied from $128$ to $2048$, and periodic boundary conditions were applied to the lattice along both the $x$ and $y$ directions. Particular attention was paid to the situation with $L=256$. The systems under consideration exhibited the rich non-equilibrium patterning, typical for materials composed of shape-anisotropic
particles~\cite{Boerzsoenyi2013}. The most striking was the formation of non-equilibrium diagonal stripe domains for aspect ratios above a specific critical value, $k\geq6$.
For shorter $k$-mers ($k\leq 5$), the diagonal patterns did not ever occur for any of the initial distributions of $k$-mers used.
Moreover, if the initial distribution of $k$-mers corresponded  to diagonal stripe domains, it was unstable and the diffusion destroyed this diagonal configuration. As at result, the final distribution was homogeneous.
Similar changes of deposition-evaporation driven self-assembly of $k$-mers on the square lattice were revealed, where the nematic order can only be stabilized for sufficiently long $k$-mers ($k\geq 7$)~\cite{Ghosh2007}.
The different self-assembly of elongated particles with dependence on their aspect ratio has also been  observed in continuous systems~\cite{Cuesta1990}. We have observed very intriguing behavior reflecting the impact of the system size on non-equilibrium self-assembly. For large scale systems with initial random placement of the $k_x$-mers and $k_y$-mers over the whole lattice, the diagonal stripe domains became less ideal and contained more defects even for $k\geq6$.

Moreover, different non-equilibrium steady states can be realized with dependence on the initial configurations. Different initial configurations were used for the placement of $k$-mers onto the lattice. For the initial r configuration, $k_x$-mers and $k_y$-mers were randomly placed  on the whole lattice. For the initial v configuration and d configuration, $k_x$-mers and $k_y$-mers were randomly placed inside the two different vertically or diagonally oriented stripes, respectively.
For the initial v configuration, the transformation of the vertical stripe domains into diagonal stripe domains was also observed when $k\geq 6$. For shorter $k$-mers ($k\leq 5$) the diagonal stripe domains were not stable for all tested initial configurations.
Although qualitatively similar effects were observed for different values of $L$ ($L=128-2048$), in the large scale systems, the diagonal stripe domains became more defected.

Finally, in the lattice adsorption of $k$-mers, the observed effects surely reflect the competition between different factors related to the discrete nature of the rods, the limited numbers of their possible orientations and the finite sample size. The relative importance of these factors still remains an open question.

\section*{Acknowledgements}
We acknowledge the funding from the National Academy of Sciences of Ukraine, Project No.~43/17-H (N.I.L., N.V.V., and V.A.G.) and the Ministry of Education and Science of the Russian Federation, Project No.~3.959.2017 (Yu.Yu.T.).

\appendix

\section{Algorithm\label{app:alg}}
%%%%%%%%%%%%%%%%%%%%%%%%%%%%%%%%%%%%%%%%%%%%%%%%%%%%%%%%%%%%%%%%%%%%%%%%%%%%
\begin{algorithmic}[1]
\STATE{}
\COMMENT{$N_{MC}$ is the total number of Monte Carlo steps}
\STATE{}
\COMMENT{$N_k$ is the number of $k$-mers to be shifted}
\FOR{$i = 1$ to $N_{MC}$}
\STATE{}
\COMMENT{One Monte Carlo step}
\FOR{$j = 1$ to $N_k$}
\STATE{Randomly select a $k$-mer}
\STATE{Randomly select a shift direction}
\STATE{Try to shift the $k$-mer in the chosen direction by one lattice site}
\STATE{}
\COMMENT{$k$-mers cannot pass through each other.}
\ENDFOR
\COMMENT{ $j$ loop}
\ENDFOR
\COMMENT{ $i$ loop}
\end{algorithmic}
%%%%%%%%%%%%%%%%%%%%%%%%%%%%%%%%%%%%%%%%%%%%%%%%%%%%%%%%%%%%%%%%%%%%%%%%%%%%

\bibliography{Diffusion}

\end{document}